\documentclass{emulateapj}
\usepackage{epsfig,lscape,natbib}
\def\lap{\lower.5ex\hbox{$\; \buildrel < \over \sim \;$}}
\def\gap{\lower.5ex\hbox{$\; \buildrel > \over \sim \;$}}

\def\ergcm2s{${\rm erg\ cm^{-2}\ s^{-1}}$}

\def\ergscm2s{${\rm erg\ cm^{-2}\  s^{-1}}$}

\def\cm-2{${\rm cm^{-2}}$}

\begin{document}

\title{The ACS Nearby Galaxy Survey Treasury VI. The Ancient Star Forming disk of NGC 404}

\author{Benjamin F. Williams\altaffilmark{1},
Julianne J. Dalcanton\altaffilmark{1},
Karoline M. Gilbert\altaffilmark{1},
Adrienne Stilp\altaffilmark{1},
Andrew Dolphin\altaffilmark{2},
Anil C. Seth\altaffilmark{3},
Daniel Weisz\altaffilmark{4},
Evan Skillman\altaffilmark{4}
}

\altaffiltext{1}{Department of Astronomy, Box 351580, University of Washington, Seattle, WA 98195; ben@astro.washington.edu; jd@astro.washington.edu; stephanie@astro.washington.edu; roskar@astro.washington.edu}
\altaffiltext{2}{Raytheon, 1151 E. Hermans Road, Tucson, AZ 85706; dolphin@raytheon.com}
\altaffiltext{3}{CfA Fellow, Harvard-Smithsonian Center for Astrophysics, 60 Garden Street, Cambridge, MA 02138; aseth@cfa.harvard.edu}
\altaffiltext{4}{Department of Astronomy, University of Minnesota, 116 Church
St. SE, Minneapolis, MN 55455; dweisz@astro.umn.edu; skillman@astro.umn.edu}


\begin{abstract} 
We present HST/WFPC2 observations across the disk of the nearby
isolated dwarf S0 galaxy NGC~404, which hosts an extended gas
disk. The locations of our fields contain a roughly equal mixture of
bulge and disk stars.  All of our resolved stellar photometry reaches
$m_{F814W}=26$ ($M_{F814W}=-1.4$), which covers 2.5 magnitudes of the
red giant branch and main sequence stars with ages $<$300 Myr.  Our
deepest field reaches $m_{F814W}=27.2$ ($M_{F814W}=-0.2$), sufficient
to resolve the red clump and main-sequence stars with ages $<$500~Myr.
Although we detect trace amounts of star formation at times more
recent than 10 Gyr for all fields, the proportion of red giant stars
to asymptotic giants and main sequence stars suggests that the disk is
dominated by an ancient ($>$10~Gyr) population. Detailed modeling of
the color-magnitude diagram suggests that $\sim$70\% of the stellar
mass in the NGC~404 disk formed by $z\sim2$ (10 Gyr ago) and at least
$\sim$90\% formed prior to $z\sim1$ (8 Gyr ago).  These results
indicate that the stellar populations of the NGC~404 disk are on
average significantly older than those of other nearby disk galaxies,
suggesting that early and late type disks may have different long-term
evolutionary histories, not simply differences in their recent star
formation rates.  Comparisons of the spatial distribution of the young
stellar mass and FUV emission in GALEX images show that the brightest
FUV regions contain the youngest stars, but that some young stars
($<$160~Myr) lie outside of these regions.  FUV luminosity appears to
be strongly affected by both age and stellar mass within individual
regions.  Finally, we use our measurements to infer the relationship
between the star formation rate and the gas density of the disk at
previous epochs. We find that most of the history of the NGC~404 disk
is consistent with star formation that has decreased with the gas
density according to the Schmidt law.  However, $\sim$0.5--1 Gyr ago,
the star formation rate was unusually low for the inferred gas
density, consistent with the possibility that there was a gas
accretion event that reignited star formation $\sim$0.5 Gyr ago.  Such
an event could explain why this S0 galaxy hosts an extended gas disk.
\end{abstract}

\section{Introduction}

While both redshift surveys and stellar population studies agree that
late-type disks form the majority of their stars by $z\sim1$
\citep[e.g.,][]{ravindranath2004,barden2005,melbourne2007,williams2008},
the ages of early-type disks are not as well constrained.  The
theoretical and high redshift observational data suggest that the old
populations of early type disks may be systematically older than those
of later type disks.  On the other hand, the morphological differences
between early and late type disks may be of recent origin, due only to
differences in star formation over the past few hundred Myr.  Such a
scenario has has been suggested for the S0 galaxy NGC~5102
\citep{davidge2008}, as well as for the differences between dEs and
dIrrs.

Unfortunately, little is known about the evolutionary history of S0
galaxies, especially in the field.  Although typical S0 galaxies
appear to differ significantly from E galaxies, it is not yet clear
where S0 galaxies fit into the Hubble tuning fork
\citep{vandenbergh2009} or what amount of star formation they
typically contain \citep{temi2008}.  While non-interacting S0
galaxies tend to have disks similar in thickness to those of late-type
galaxies \citep{pohlen2004,li2009}, other properties of S0 disks
differ significantly from those of late-type galaxies.  For example,
their color gradients are less pronounced than those of late-type
disks \citep{macarthur2004}, and S0 galaxies are typically fainter
than later type disks \citep{vandenbergh2009}. S0 galaxies are
predominantly found in galaxy clusters and are mostly stripped of {\sc
H i} \citep{poggianti2009,vandenbergh2009}, whereas field S0 galaxies
like NGC~404 are less common and could be a type of post-starburst
dwarf \citep{davidge2008}.

As the nearest face-on example of an S0 galaxy \citep[3.05~Mpc,
  m-M$_0$ = $27.42\pm0.02$; M$_K$ = -18.9; $A_V$=0.08;
  $i$=11$^{\circ}$;
  $V_{c}=190$~km~s$^{-1}$][]{dalcanton2009,schlegel1998,delrio2004},
  NGC~404 has been previously studied in some detail. Although its
  optical morphology suggests that NGC~404 is a classic ``red and
  dead'' early-type galaxy, the center appears to have a significant
  young component.  With long-slit spectroscopy of the galaxy core,
  \citet{bouchard2010} found evidence for intermediate-aged, and old
  stellar populations, along with a much younger component that fades
  very quickly with radius.  The nucleus also has a dominant 1 Gyr old
  population \citep[$\sim$50\% of the mass,][]{seth2010}.  UV
  spectroscopy also indicates some very young ($\leq$ 10 Myr) stars in
  the galaxy center \citep{maoz1998}. Farther out, NGC~404 has a very
  large ($>$20 kpc) stellar disk that is known to be dominated by red
  giants \citep{tikhonov2003}.  This disk also contains a large amount
  of {\sc H i} gas in a well-defined inner disk ($<$5 kpc) and a
  warped outer disk \citep{delrio2004}, although the density of gas is
  low ($<$2$\times$10$^{20}$ M$_{\odot}$ pc$^{-2}$ throughout) and
  only a faint ring of star formation is detected in the far
  ultra-violet
  (FUV)\footnote{http://www.galex.caltech.edu/newsroom/glx2008-02r.html}
  \citep{wiklind1990}.

The origin of the gas disk of NGC~404 is somewhat
mysterious. \citet{delrio2004} suggest that the disk is likely due to
a recent (0.5--1 Gyr ago) merger event with a dIrr galaxy, motivated
by their detection of a possible warp in the disk.  However, NGC~404
is very isolated for a galaxy of its mass \citep[$\Theta$=-1.0;
see][for details]{karachentsev2004}. With no other galaxies of any
kind detected within 1~Mpc of NGC~404, such a merger is not obviously
consistent with the galaxy's environment.  The gas disk could
therefore simply be due to late time gas infall from filaments rather
than accretion of bound objects.  Furthermore, a gas disk of such low
density could have a very long lifetime given the low star formation
rate, suggesting that it could have been accreted far earlier. By
comparing the stellar populations and gas densities, we can constrain
the origin of this unusual disk.

We examine the stellar populations of NGC~404 in detail using deep
observations from HST.  We determine the star formation history (SFH)
of several portions of the galaxy by fitting the distribution of stars
in color-magnitude diagrams (CMDs) with model distributions determined
from stellar evolution isochrones \citep[as described
in][]{dolphin2002}.  Section 2 describes our data set and analysis
procedures.  Section 3 presents the results of our measurements.
Section 4 interprets the measurements in the context of the gas
properties and our understanding of S0 galaxies, and Section 5
summarizes our conclusions.  We assume a distance of 3.05~Mpc
\citep{dalcanton2009} for conversions of angular measurements to
physical distances and adopt an inclination angle $i$=11$^{\circ}$
\citep{delrio2004} for surface density measurements.  We adopt a
five-year WMAP \citep{dunkley2009} cosmology for all conversions
between time and redshift.

\section{Data Acquisition, Reduction, and Analysis}

\subsection{Acquisition}\label{acquisition}

From 2007-Aug-08 to 2007-Sep-20, we observed a field in the NGC~404
disk located at R.A.~(2000) = 17.32325 (01:09:17.6), decl.~(2000) =
35.74856 (+35:44:55) with a rotation angle PA\_V3=50.0 degrees.  From
2009-Feb-16 to 2009-Feb-20, we performed shallower observations for 2
fields located at R.A.~(2000) =17.36697 (01:09:28.1), decl.~(2000) =
35.76117 (+35:45:40) with a rotation angle PA\_V3=230.0 and
R.A.~(2000) =17.333368 (01:09:20.0), decl.~(2000) = 35.70205
(+35:42:07) with a rotation angle PA\_V3=230.0. Figure~\ref{field_loc}
shows outlines of the fields' locations.  Our field locations were
chosen to maximize the number of disk stars and avoid crowding.

In the deep field, we obtained 15 full-orbit exposures with the WFPC2
\citep{ford1998} through the F606W (wide $V$) filter, and 29
full-orbit exposures through the F814W ($I$ equivalent) filter.  These
data totaled 39000~s and 75400~s of exposure time in F606W and F814W,
respectively.  In the other two fields, we obtained 2 orbits through
F606W, totaling 4800~s, and 4 orbits through F814W, totaling
9600~s. All images were calibrated in the HST pipeline with CALWP2
using OPUS version 2006\_6a for the 2007 data and 2008\_5c for the
2009 data.

To expand our radial coverage, we also reduced 2 fields in the outer
disk, previously studied and named S2 and S3 \citep[see][for more
  details]{tikhonov2003}.  These fields lie $\sim$8$'$ (7 kpc) SW of
the nucleus and were taken as part of GO-5369.  They contain 1200 sec
of exposure in F606W each.  Field S3 contains 4200 sec of exposure in
F814W, while S2 contains only 2100 sec in F814W.

\subsection{Reduction}\label{reduction}

The data reduction and photometry for the ANGST survey are fully
described in \citet{dalcanton2009}.  For convenience we provide a
brief summary of the techniques here.  

The photometry was measured simultaneously for all of the objects in
the uncombined images using the software package HSTPHOT
\citep{dolphin2000}.  This package is optimized for measuring
photometry of stars on WFPC2 images using the well-characterized and
stable point spread function (PSF) calculated with
TinyTim.\footnote{http://www.stsci.edu/software/tinytim/} The software
fits the PSF to all of the stars in each individual frame to find PSF
magnitudes.  It then determines and applies the aperture correction
for each image using the most isolated stars, corrects for the charge
transfer efficiency of the WFPC2 detector\footnote{July, 2008 formulae\\
http://purcell.as.arizona.edu/wfpc2\_calib/2008\_07\_19.html},
combines the results from the individual exposures, and converts the
measured count rates to the VEGAmag system.

\begin{figure}
\centerline{\psfig{file=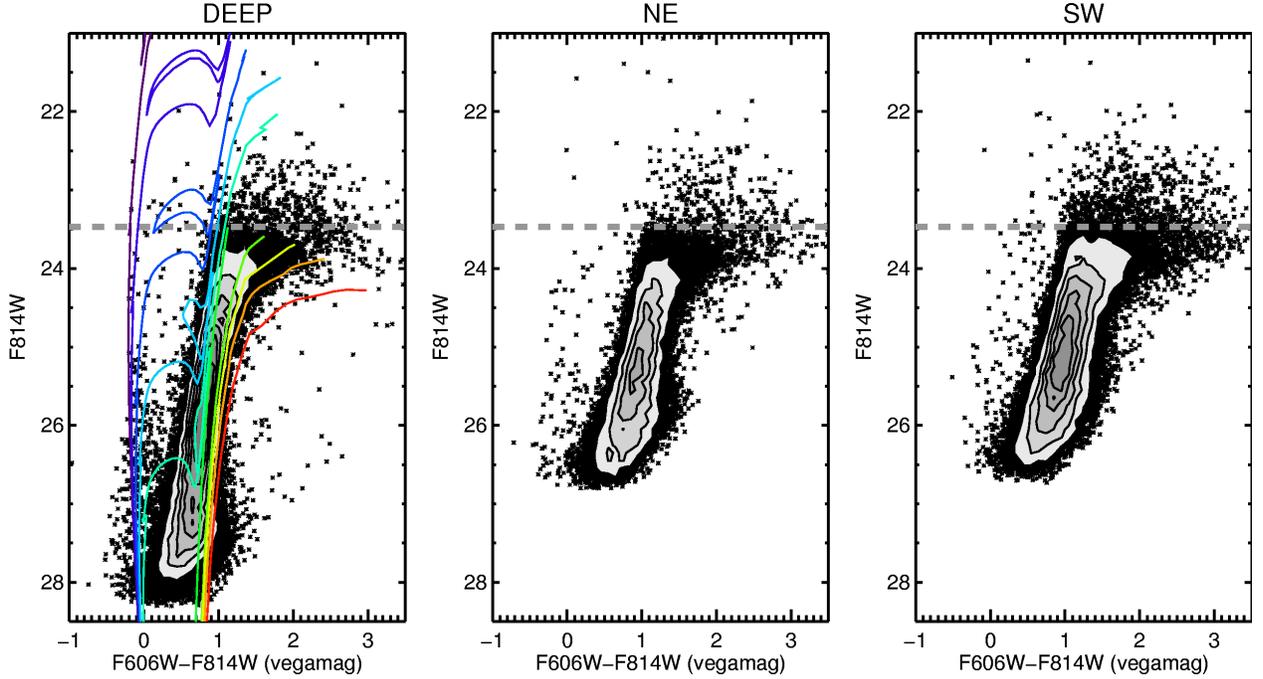,width=6.5in,angle=0}}
\caption{The color-magnitude diagrams of our 3 WFPC2 fields
(\S~\ref{reduction}). Contours denote the density of points in areas
of the plot that would otherwise be saturated.  Overlaid on the deep
field CMD are example isochrones from the \citet{girardi2002} models
shifted to $m-M_0=27.48, A_V=0.08$ (from blue to red: [M/H]=-0.4 and
log(age)~=~7.3,7.6,8.0,8.3,8.6, followed by log(age)=10.0 and
[M/H]~=~-1.3,-0.7,-0.4,-0.2,0.0, respectively).  A gray dashed line
marks the tip of the red giant branch as measured by
\citep{dalcanton2009}.}
\label{cmds}
\end{figure}

\begin{figure}
\centerline{\psfig{file=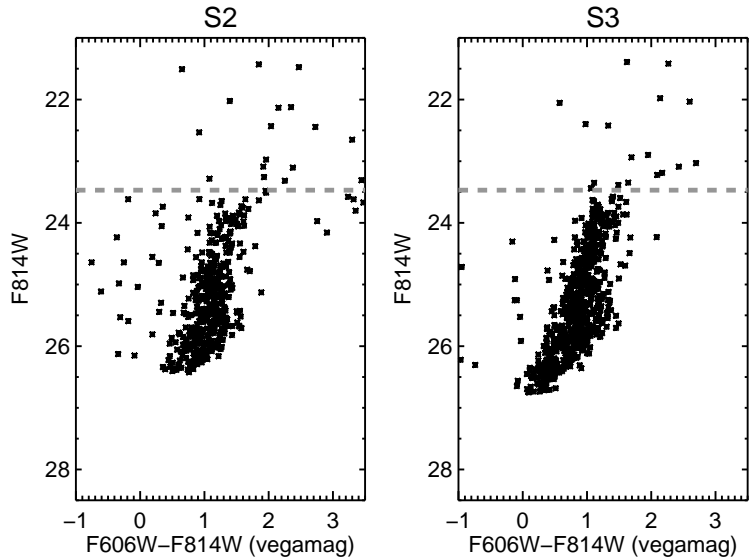,width=4.0in,angle=0}}
\caption{The color-magnitude diagrams of the 2 archival WFPC2 fields
in the outer disk (\S~\ref{reduction}) previously studied by
\citet{tikhonov2003}.  A gray dashed line marks the tip of the red
giant branch as measured by \citep{dalcanton2009}.  We note the TRGB
appears slightly fainter in these fields, explaining why the distance
measurement of \citet{tikhonov2003} was greater than that of
\citet{dalcanton2009}.}
\label{outer_cmds}
\end{figure}

The HSTPHOT output was then filtered to only allow objects classified
as stars with signal-to-noise (total counts from the star to total
noise) $>$4 in both filters.  The list was further culled using
sharpness ($|F606W_{sharp} + F814W_{sharp}| < 0.27$) and crowding
($F606W_{crowd} + F814W_{crowd} < 0.7$).  The sharpness cut was chosen
based on the distribution of values in the original catalog.  The
crowding parameter gives the difference between the magnitude of a
star measured before and after subtracting the neighboring stars in
the image.  When this value is large, it suggests that the star's
photometry was significantly affected by crowding, and we therefore
exclude it from our catalog.  Quality cuts based on the $\chi$ values
were also considered, but they were rejected when a correlation was
found between $\chi$ and the local background.  Our final star
catalogs contained 40793, 22332, and 33365 stars for the deep, NE, and
SW fields, respectively.  The archival outer disk fields S2 and S3 are
much more sparsely populated.  We obtained reliable photometry for
only 468 and 692 stars for S2 and S3, respectively. The final CMDs for
the fields are shown in Figures~\ref{cmds} and \ref{outer_cmds}.

\subsection{CMD Fitting}\label{fitting}

We measured the star formation rate and metallicity as a function of
stellar age using the software package MATCH \citep{dolphin2002}.  We
fit the observed CMDs (with magnitude cuts set to limits provided in
Table~\ref{table}) by populating the stellar evolution models of
\citet[][with updates in \citealp{marigo2008}]{girardi2002} with a
\citet{salpeter1955} initial mass function (IMF) for a grid of assumed
distance and foreground extinction values to allow for systematic
differences in stellar evolution models and/or systematic photometric
errors.  The choices of software and models used for the ANGST project
are discussed in detail in \citet{williams2008} and summarized in
\citet{dalcanton2009}.

\begin{figure}
\centerline{\psfig{file=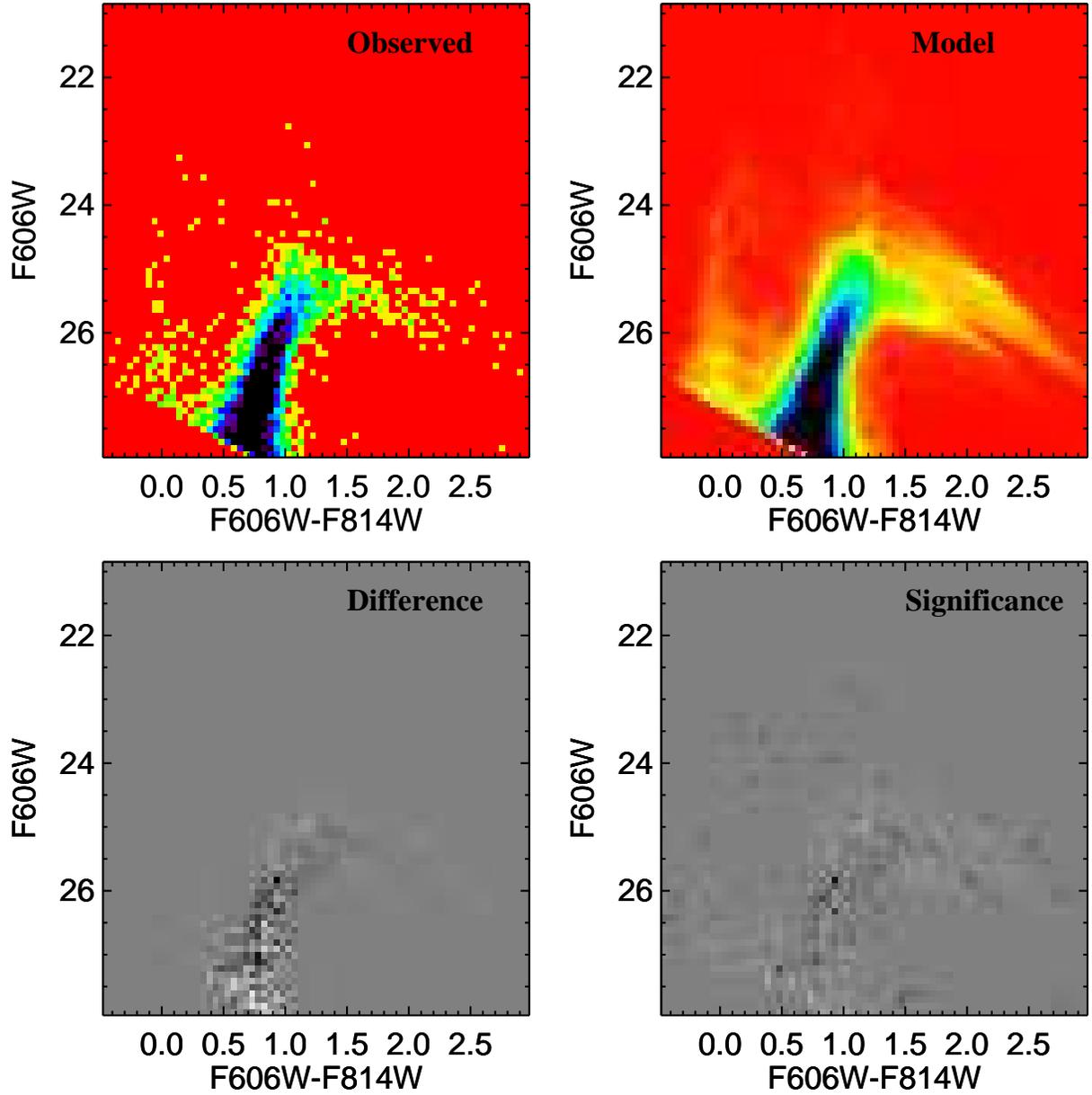,width=6.5in,angle=0}}
\caption{Our best-fit model CMD to the outermost annulus of the deep
field data set (\S~\ref{fitting}).  This fit is our most reliable as
it comes from our deepest photometry.{\it Upper Left:} Our observed
CMD cut at the magnitude limits used for the fit. {\it Upper Right:}
The best-fit model CMD.  {\it Lower Left:} The difference between the
data and the model.  Darker colors denote excess stars in the data.
Whiter colors denote excess stars in the model.  The range is from -19
(lightest) to +23 (darkest), corresponding to fractional errors of
$\pm$0.5.{\it Lower Right:} The statistical significance of the
residuals shown in Lower Left.  The range is -5$\sigma$ (lightest) to
+10$\sigma$ (darkest).
}
\label{residuals}
\end{figure}

\begin{figure}
\centerline{\psfig{file=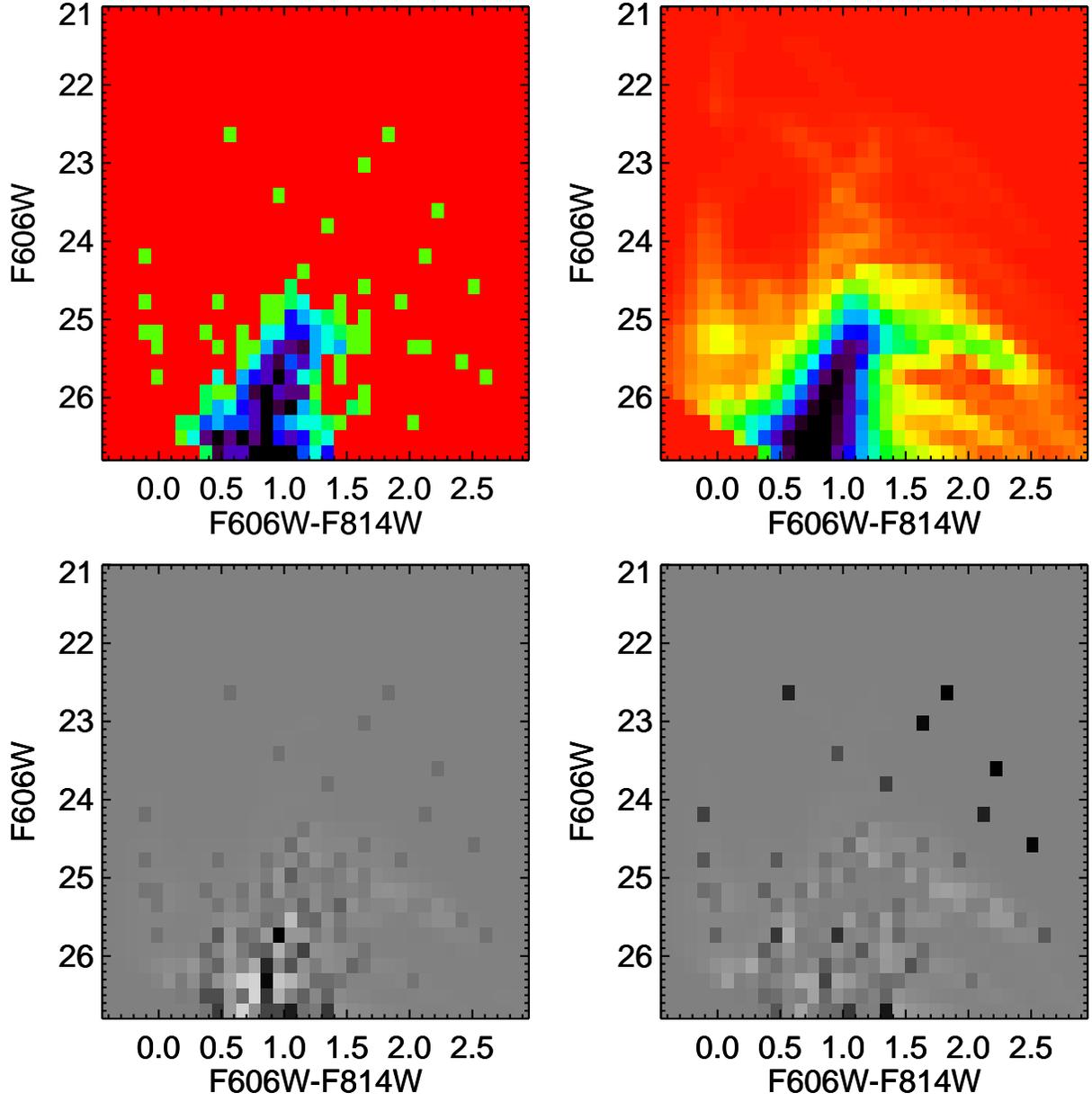,width=6.5in,angle=0}}
\caption{Our best-fit model CMD to the deeper of the two outer fields
(S3). Panels are the same as Figure~\ref{residuals}.  The range is
from -5 (lightest) to +8 (darkest), corresponding to fractional errors
of $^{+0.4}_{-0.6}$.{\it Lower Right:} The statistical significance of
the residuals shown in Lower Left.  The range is -1.5$\sigma$
(lightest) to +4.2$\sigma$ (darkest).  }
\label{s3_fit}
\end{figure}

The best fits provide the combination of ages and metallicities that
are contained in the observed field.  We attempted to fit the data
with a spread in the model photometry along the reddening line to
account for the effects of differential reddening. However, applying a
spread in reddening of $A_V=0.5$ to the models degraded the quality of
the CMD fits, showing that differential reddening does not
significantly affect our measurements in NGC~404. The data from the
deep field were best fit by a single foreground reddening
$A_V$=0.1$\pm$0.06 and $m-M_0$=27.48$\pm$0.09 (see
Figure~\ref{residuals}). This distance modulus is consistent with, but
larger than, the value measured by the ANGST survey
\citep[$m-M_0$=27.42$\pm$0.02][]{dalcanton2009}.  The best fit values
compensate for any systematic differences between the data and overall
model isochrones, whereas the survey value isolated the
well-determined location of the tip of the red giant branch in order
to measure the best distance. Since these distance and extinction
values provided the best overall fit of the models to the data, we
performed our fits to the data of every region assuming these values.
Our uncertainties in star formation rate account for changes in the
SFH measured if the assumed value for the distance modulus was
$\pm$0.15 mag away from the chosen value and if the extinction value
was $\pm$0.1 mag away from the chosen value.  We note that in
Figure~\ref{outer_cmds} the apparent magnitude of the tip of the RGB
appears fainter than in our more populous CMDs.  Our fits to these
CMDs (Figure~\ref{s3_fit}) were not significantly improved by allowing
a greater distance modulus, indicating that the small numbers of stars
in these fields cause the tip of the RGB to be under-populated.  On
the other hand, the SFHs of these fields could be significantly
different than the inner fields. The shallow depth and low numbers of
stars in these fields limits our ability to constrain the age. Our
fits show only that 99\% of the stars are older than 1.6 Gyr. Thus, it
is possible that most of these stars are only a few Gyr old, which
could produce a fainter TRGB. Unfortunately, there is not enough data
to constrain whether the cause is undersampling or age. The CMD can be
fitted equivalently well by either possibility.

Systematic errors are determined by the MATCH package by comparing the
results of SFHs from fits to the data with different values for the
distance and foreground reddening to the field.  These errors are then
added in quadrature to the random errors governed by our sampling of
the CMD.  The random errors are determined by randomly drawing from
the observed CMD to produce Hess diagrams that vary due to the Poisson
statistics of our photometric sample.  By producing and fitting 100 of
these Monte Carlo CMDs, we are able to determine the rms of the
residuals between SFHs from these fits and those from the fits to the
original data.  The combined 1$\sigma$ error measurements therefore
account for the uncertainties in the distance to the galaxy, the
foreground reddening, any systematic shifts between the model colors
and magnitudes and our measured photometry, as well as the number of
stars and features present in our CMD.

Our Monte Carlo tests are also used to determine our time sensitivity
\citep[i.e., the dependence of our SFH on our choice of time binning;
  see][for a full description of the
  technique]{williams2009b}. Briefly, we calculate the standard
deviation of the maximum likelihood value from our 100 runs.  We
assume that any fit to the data more than one standard deviation away
from the best fit is unacceptable.  We then rerun our fits while
suppressing star formation in various time bins.  If the fit quality
does not change significantly (by more than one standard deviation),
we continue to expand the length of these removed time bins until the
software can no longer find an acceptable fit.  At this time
resolution, we can be confident that our data provide meaningful
constraints on the SFH.  The final time bins are all sensitive enough
that their removal from the SFH results in an unacceptable CMD fit.

\subsection{Depth}\label{depth}

Photometric depth determines the precision with which we can recover
the SFH of a region. The effects of age and metallicity are more
difficult to distinguish with shallow photometry than with deep
photometry.  Our deepest photometry comes from our deep field and
reaches the red clump in the least crowded region.  This photometry
therefore provides the most leverage for breaking the degeneracy
between the age and metallicity of the old populations. In contrast,
the data from our 2 shallow fields provide the least of this leverage.
However, since the stellar populations should be well mixed at ages
$\gg$1 Gyr, we used the metallicity distribution for the old stars
($>$2~Gyr) as determined from the fit to our full deep field to limit
the range of allowed metallicities at each age in the fits to the
shallower data.  When the free parameters used to fit the shallower
data were limited, we found the resulting age distribution of the
ancient populations of the shallower fields to be consistent with
those of the deep data, and the quality of the fit remained in the
acceptable range (within 1 standard deviation of the value obtained
when the full grid of free parameters was allowed). In what follows,
the full-field SFHs for the shallower fields are the best fits
possible with the restriction that the ancient population ($>$2~Gyr)
contain only the metallicities at each age that contributed to the
best fit of the deep full-field data.

\subsection{Field Division}\label{division}

\begin{figure}
\centerline{\psfig{file=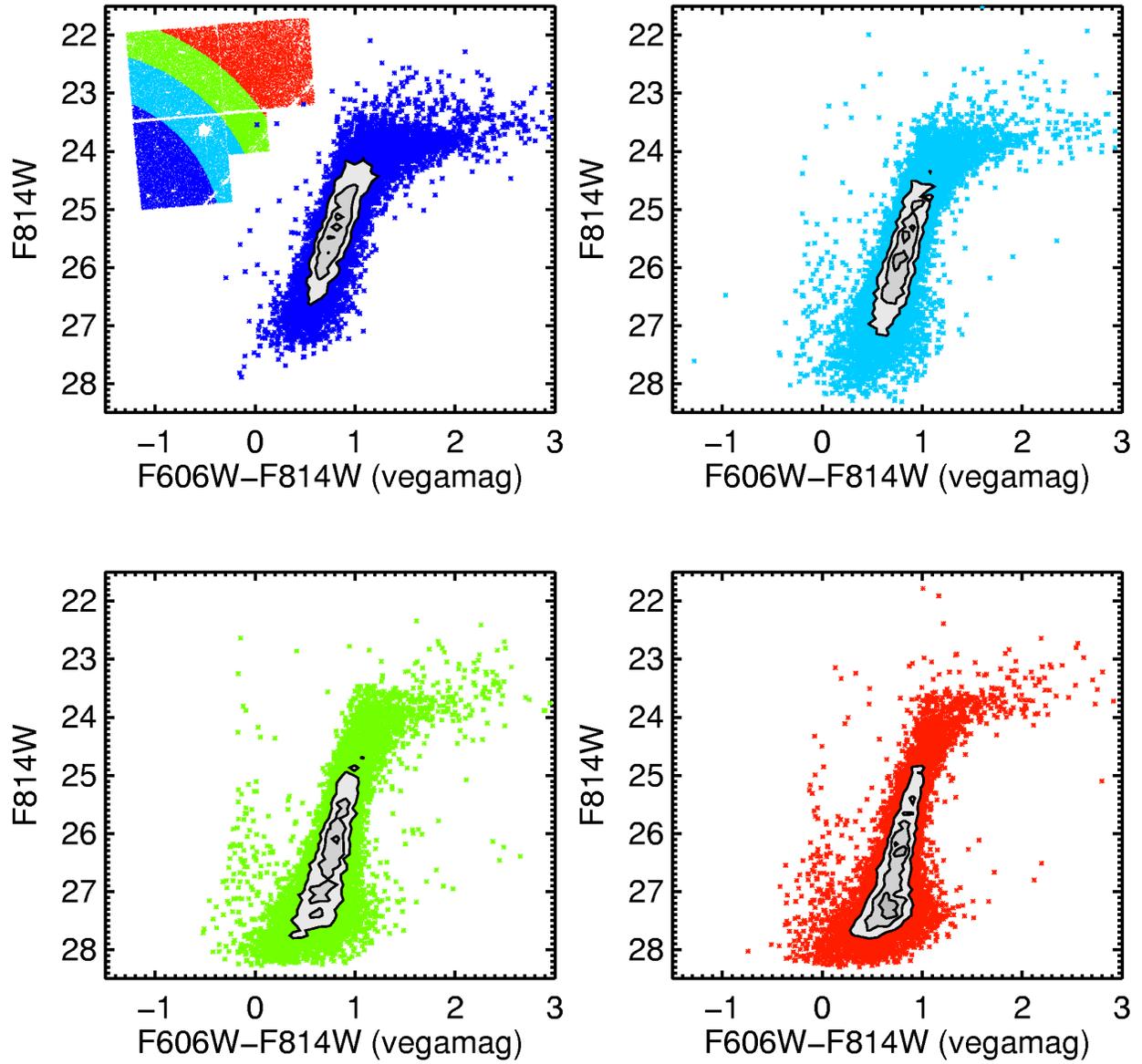,width=6.5in,angle=0}}
\caption{The CMDs of the 4 annuli we defined for studying radial
variations in the stellar populations (\S~\ref{division}). The inset
of the upper-left CMD shows the locations of the stars in each annulus
on our deep WFPC2 field.  Annuli were chosen to each contain
$\sim$10000 stars in the deep field.  Radii of the boundaries are
64$''$, 124.0$''$, 160.0$''$, 190.0$''$, and 300.0$''$.}
\label{annuli_cmds}
\end{figure}

In addition to the full field CMDs, we divided our inner disk fields
into 4 annuli to look for radial trends of the stellar populations and
to help minimize differential crowding effects across each field.  The
annuli were chosen so that each region contained $\sim$10000 stars in
our deep field data.  Their parameters of the annuli are provided in
Table~\ref{table}.  Figure~\ref{annuli_cmds} shows the CMDs of the
different annuli in our deep field.  These CMDs show clearly that
crowding in the inner regions is significantly worse than in the outer
regions of our fields (as shown by the limited depth of photometry in
the inner fields).  In order to constrain the old populations of the 2
inner annuli, we limited the number of free parameters by allowing
only the metallicities that provided the best fit to the outermost
annulus of the deep field for ages $>$2~Gyr. This restriction produced
a acceptable fits to the CMDs while forcing the ancient population of
the innermost regions to be of similar metallicity to that of the
outermost region.

\subsection{Spheroid Contamination}\label{contamination}

\citet{baggett1998} performed an early bulge-disk decomposition on
NGC~404 from digitized photographic plate observations in the
$V$-band, finding a bulge effective radius of 63.8$''$ and a disk
scale length of 129.5$''$.  Their parameters indicate that the disk
population is responsible for 50\% of the galaxy light at
$\sim$165$''$ (near the inner boundary of our third annulus from the
center).  Adopting this decomposition, our data are centered around
the transition region between the domination of the bulge component
and the domination of the disk component, suggesting our 2 outer
annuli probe regions that are dominated by the disk
population. However, more recently, the light profile of the bulge was
measured with {\it HST} data, and the resulting effective radius was
found to be smaller \citep[38$''$,][]{seth2010}.  We performed our own
fit to the light profile using the {\it Spitzer} 3.6$\mu$m image from
the local volume legacy (LVL)
survey\footnote{http://irsa.ipac.caltech.edu/data/SPITZER/LVL/} for
the central portions and normalizing to the star counts from our {\it
HST} data.  We found a similar effective radius (42$''$).  Our fit
requires that 75\% of the stars belong to the disk at 165$''$.  These
measurements suggest that the amount of spheroid contamination
inferred from the \citet{baggett1998} decomposition is an upper-limit,
and all of our data are dominated by disk stars.

\section{Results}

\subsection{The Ancient Stellar Population of the Disk}\label{ancient}

\begin{figure}
\centerline{\psfig{file=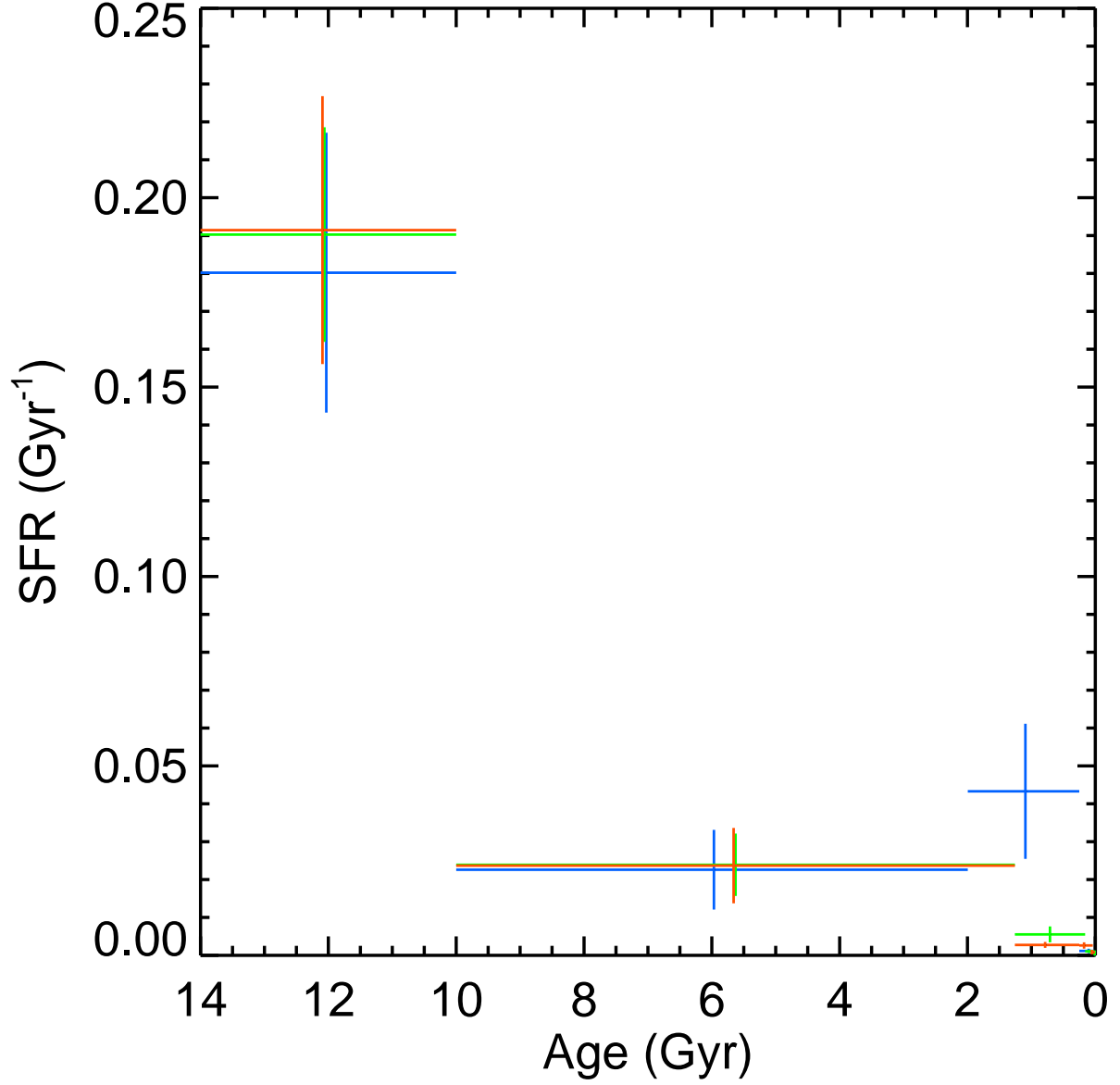,width=6.5in,angle=0}}
\caption{The SFH of our 3 fields, normalized by total stellar mass in
the field.  This normalization produces units of fraction of the total
field stellar mass produced each Gyr (Gyr$^{-1}$).  Blue, green, and
red correspond to the results from the deep, SW, and NE fields
respectively. The units can be converted to M$_{\odot}$ yr$^{-1}$
kpc$^{-2}$ by multiplying the Gyr$^{-1}$ values by 0.038, 0.080, and
0.038, respectively.  The SFHs of all fields agree and show most of
the star formation occurring before 10~Gyr ago (\S~\ref{ancient}).}
\label{sfh_full}
\end{figure}

\begin{figure}
\centerline{\psfig{file=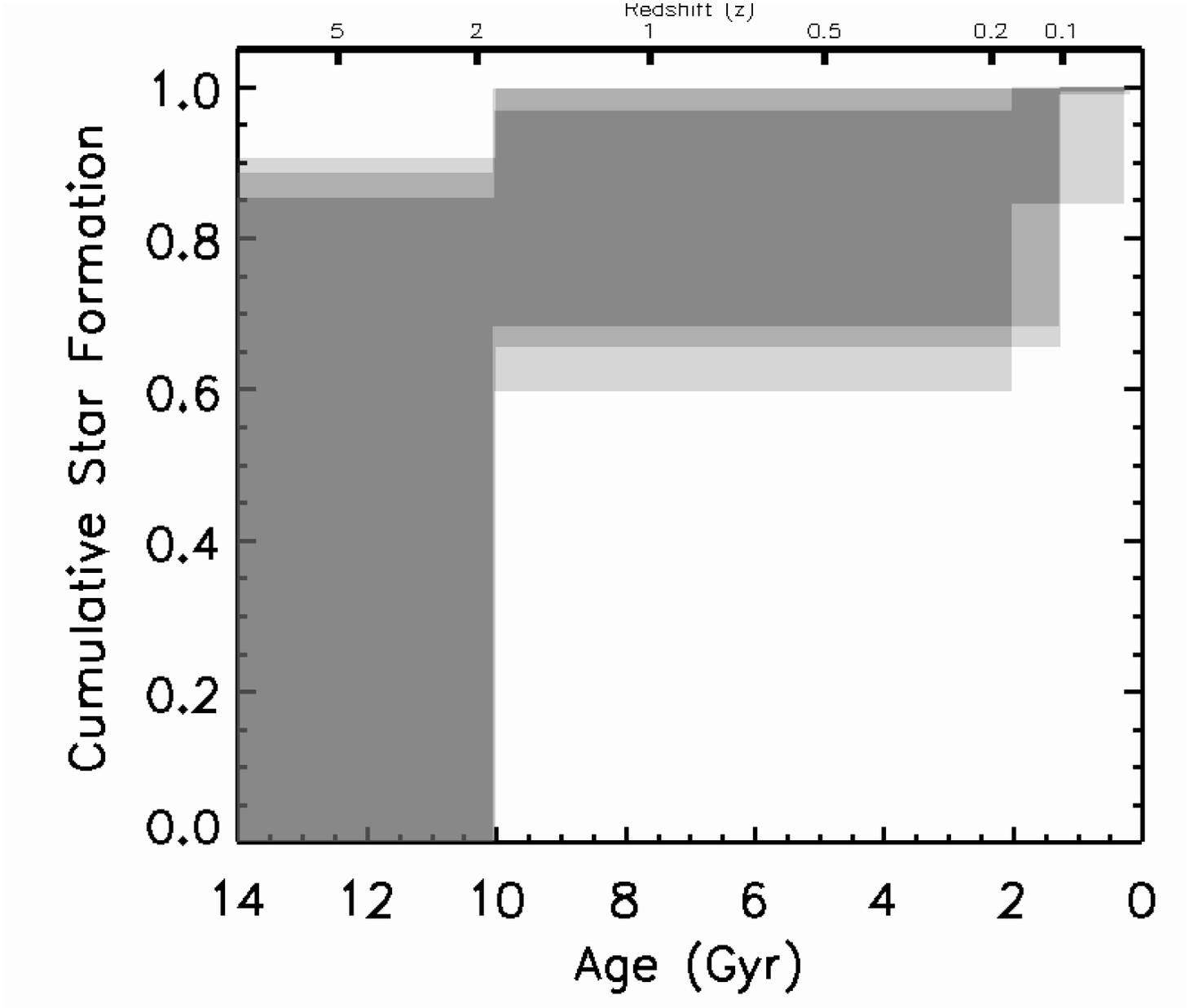,width=6.5in,angle=90}}
\caption{The cumulative SFH of our 3 fields. The results from each
individual field are shaded in light gray.  Areas of overlap between
the SFH of different fields are darker.  The darkest gray denotes the
SFH that agrees with all of the fields.  The results from all fields
overlap significantly and suggest a disk dominated by very old
($\gap$10 Gyr) stars (\S~\ref{ancient}).}
\label{cum_full}
\end{figure}

The SFHs for our 3 fields are shown in Figure~\ref{sfh_full} (SFR
versus time) and Figure~\ref{cum_full} (the cumulative fractional star
formation versus time).  The SFH of the outer disk (archival fields S2
and S3) was not well-constrained due to the small number of stars.  We
note only that the weak constraints obtained from these fields
($>$98\% of the stellar mass formed before 1.6~Gyr ago) were
consistent with the age distributions of the stars in the outer disk
being the same as those in our 3 inner disk fields.

The stellar populations are clearly dominated by stars formed prior to
10~Gyr ago in all 3 fields.  The cumulative age distribution of all of
the fields is consistent with $\gap$70\% of the stellar mass formed by
$z\sim\,2$.  Most of this constraint comes from the deep data, which
provided the metallicity constraints for the shallower fields and has
the greatest fraction of stellar mass formed prior to 10~Gyr. This
initial star formation was then followed by roughly continuous star
formation at low levels to very recent times ($\lap$10~Myr ago).
GALEX ultraviolet imaging is consistent with this low level of star
formation continuing through the past few Myr and possibly to the
present day.  The low recent star formation rate is consistent with
the low surface density of {\sc H i} \citep[$\sim$1.2
M$_{\odot}$~pc$^{-2}$,][]{delrio2004}.

\begin{figure}
\centerline{\psfig{file=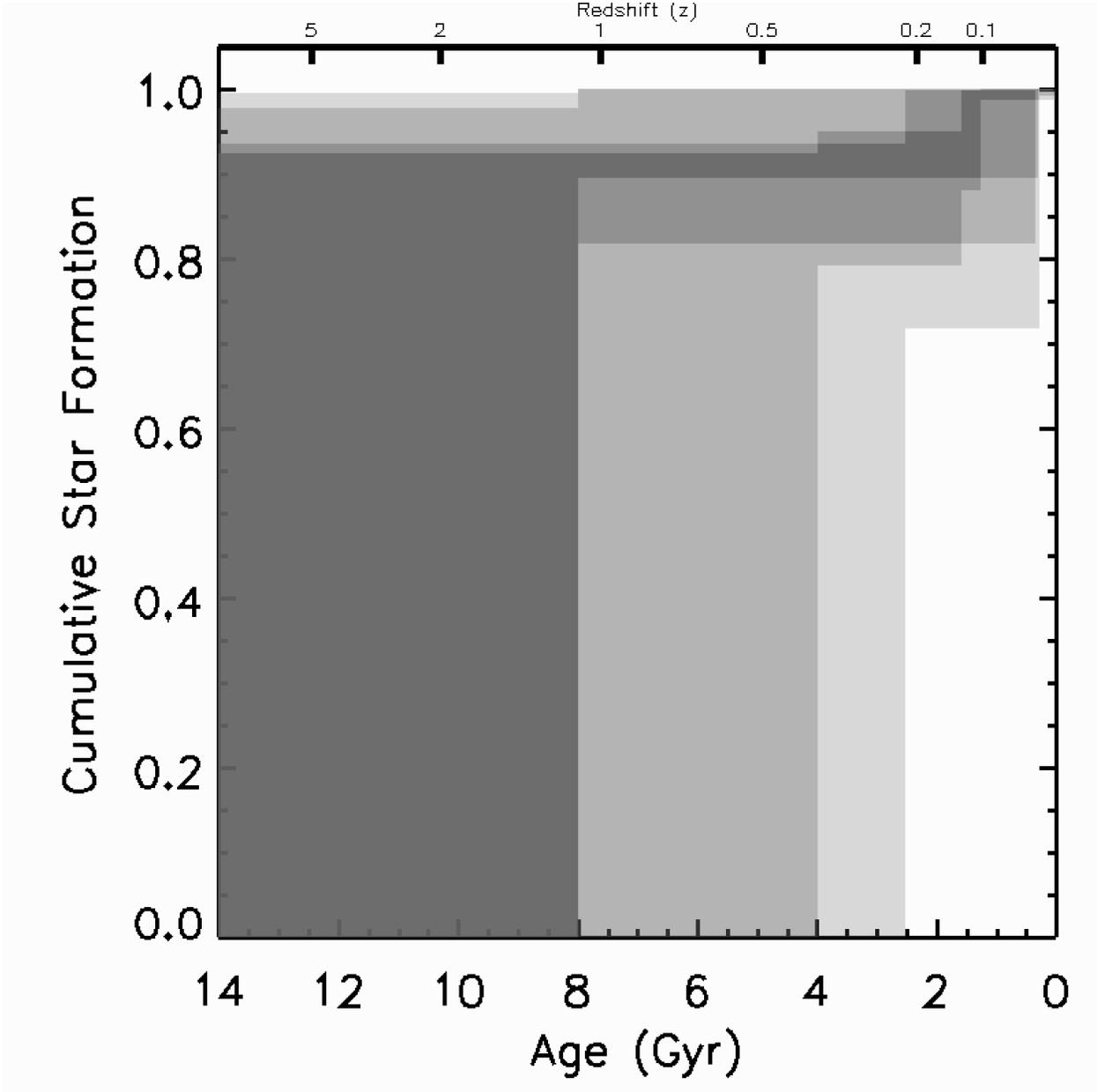,width=6.5in,angle=90}}
\caption{The cumulative star formation history of our 4 annuli in just
the deep field. The results from each individual annulus are shaded in
light gray.  Areas of overlap between the SFH of different annuli are
darker.  The darkest gray denotes the SFH that agrees with all of the
annuli.  The results from all annuli overlap significantly, although
the inner annuli do not provide as much constraint on the age
distribution of stars older than 2 Gyr as the deeper outer annuli.  The
fact that the SFHs of all annuli overlap with that of the
well-constrained outermost annulus suggests little radial variation of
the stellar populations in the deep field (\S~\ref{ancient}).}
\label{cum_ann_deep}
\end{figure}

\begin{figure}
\centerline{\psfig{file=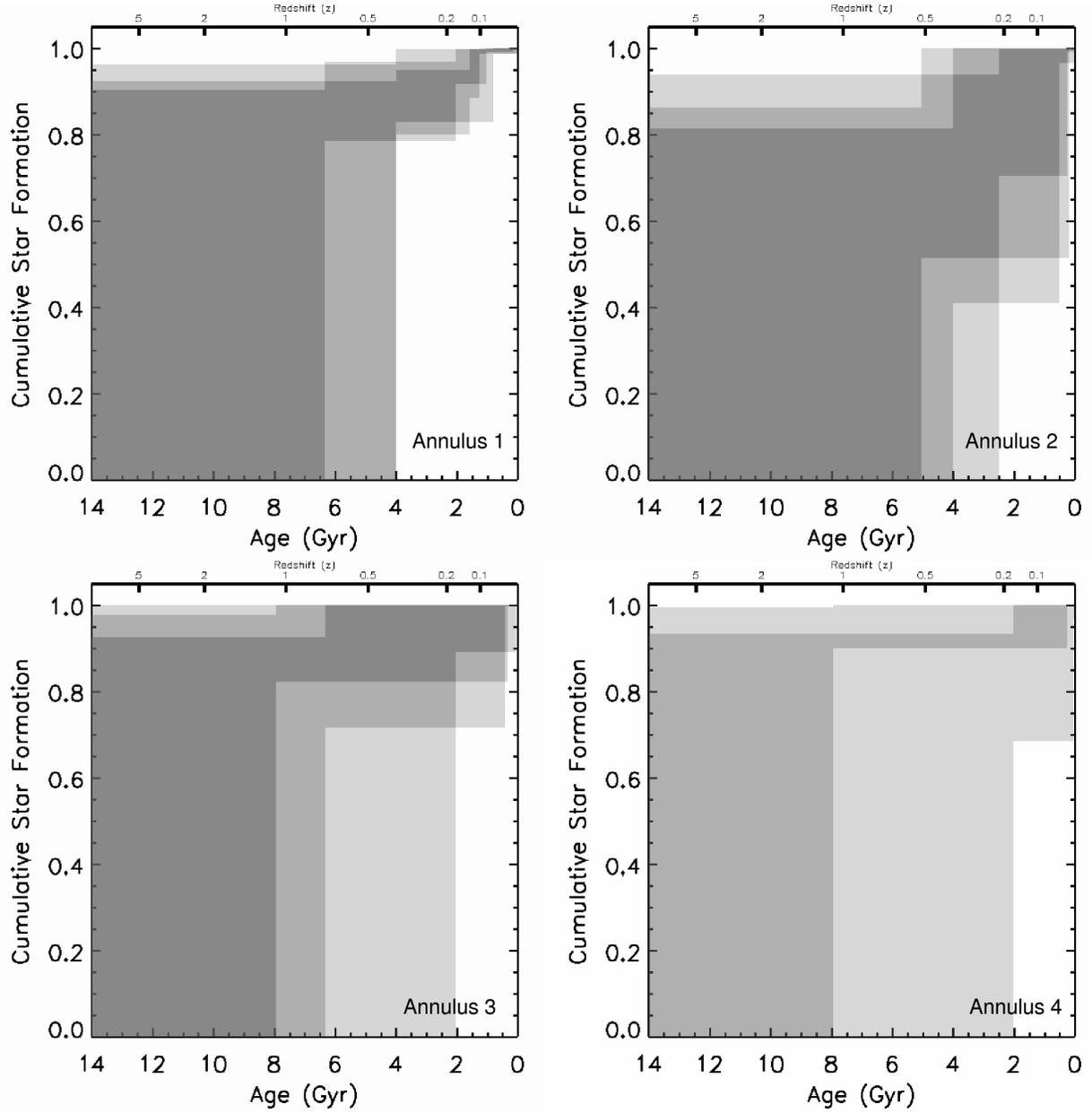,width=6.5in,angle=0}}
\caption{The cumulative star formation history of our 4 annuli. The
results from each field are shaded in light gray.  Areas of overlap
between the SFH of the same annulus in different fields are darker.
The darkest gray denotes the SFH that agrees with all of the data.
The results from all fields overlap significantly, but the deep field
constrains the stellar populations to be dominated by the oldest stars
(\S~\ref{ancient}).}
\label{cum_ann_all}
\end{figure}

Our results show that the stars in our disk fields are older than any
other disks previously-studied in this manner; however, it is not
clear if all parts of our fields are uniformly old.  To address this
question, we measured the SFH of our radial annuli.  The cumulative
SFHs of the stars in each of the 4 annuli are shown in
Figures~\ref{cum_ann_deep} and \ref{cum_ann_all}.  These SFHs do not
generally constrain the older populations as well as the full-field
fits because the CMDs contain fewer stars.  However, they are all
consistent with $\gap$70\% of the stellar mass being formed by
$z\sim\,2$.  The deepest photometry in our sample, which comes from
the outermost annulus of the deep field, constrains $\gap$90\% of the
stellar mass to be older than 8~Gyr ($z\sim$1).  The result from the
outermost annulus of the shallower NE field is consistent with that
result but does not provide nearly as tight of a constraint ($>$70\%
of the stellar mass older than 2 Gyr) due to the lack of any of the
red clump feature in the CMD.  In the end, the SFHs of our different
annuli allow the possibility that the inner annuli are a few Gyr
younger on average than the outer annuli; however, it is not clear
whether this difference is due to bulge contamination, shallower
photometry, or true differences in age.

Because the full SFH measurements in our radial annuli did not show
any conclusive differences between the stellar populations of the
NGC~404 disk with radius, we attempted a more traditional method of
comparing the AGB/RGB ratios in different
regions. \citet{tikhonov2003} found that inside of $r\sim$150$''$, the
ratio of AGB to RGB stars decreases with radius, which interpret as
evidence for most of the AGB population being associated with the
bulge.  Our constraints on the detailed age distribution in the inner
annuli allow the possibility that the inner regions are younger than
the outer regions.  Furthermore, our innermost annulus has a slightly
bluer RGB and a higher AGB/RGB ratio than the other annuli studied,
consistent with the \citet{tikhonov2003} results.  Gaussian fits to a
slice through the CMDs in Figure~\ref{annuli_cmds} at $m_{F814W} =
24.75$ give a central RGB color of $m_{F606W}-m_{F814W}$ =
0.95$\pm$0.01 in our innermost annulus, while those of the outer
annuli and outer disk fields are all $\geq$0.98.  The ratio of AGB
stars ($m_{F814W} < 23.5$) to RGB stars ($25.0 > m_{F814W} > 23.5$) is
0.056$\pm$0.005 in the innermost annulus, while those of the outer
annuli are all $<$0.05.  This ratio was not reliably constrained in
the outer disk fields (S2 and S3), due to the low numbers of stars.
Overall, the results of our full CMD fitting, as well as direct
measurements of our CMDs' features, are consistent with the suggestion
in \citet{tikhonov2003} that NGC~404 contains a bulge population that
is on average {\it younger} than the disk.  However, with the current
data we cannot distinguish between this possibility and the
possibility of radial variations in the mean age of the disk itself.

\begin{figure}
\centerline{\psfig{file=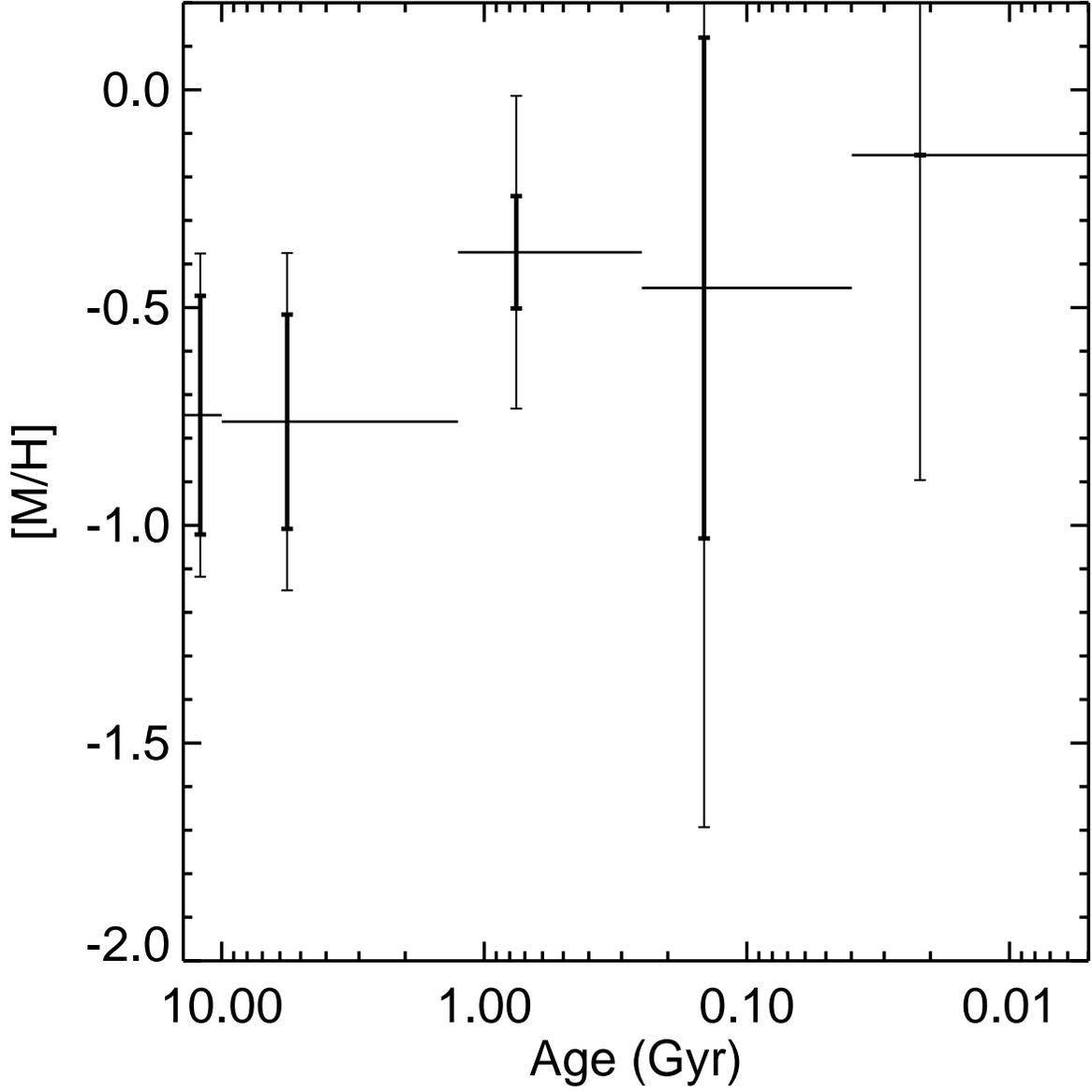,width=6.5in,angle=0}}
\caption{The chemical enrichment history of NGC~404 from our CMD fit
to the deep field. Thick error bars mark the spread in metallicity,
while thin error bars show the estimated uncertainty in the mean
metallicity and spread.  The ancient stars are dominated by
metallicities $>$-1.0, and there is evidence of an increase at times
more recent than $\sim$1~Gyr (\S~\ref{ancient}).}
\label{z_full}
\end{figure}

The metallicity history from our deep field is shown in
Figure~\ref{z_full}.  As with many other nearby galaxies, we find that
NGC~404 enriched to [M/H]$\gap$-1 quickly and then appears to have
undergone more modest subsequent chemical enrichment.  The errors are
very large on the metallicity of the young populations both because
the number of young stars is low and because the properties of the
upper main sequence is not strongly affected by metallicity.
Therefore, while our metallicity history does not show any obvious
discontinuities, if recently accreted gas was of different metallicity
than the previously-existing gas disk, it is not clear we would have
detected the difference with statistical significance.  Indeed, the
allowed metallicity range in the 30--300 Myr timebin allows any
metallicity [M/H]$\geq$-1.7.

From these results we can assign the typical age and metallicity of
the stars in these portions of NGC~404 as 12$\pm$2~Gyr and
-0.75$\pm$0.37.  These values are consistent with those determined
from fits to optical spectra of the bulge \citep[$>$50\% of the stars
$>$5 Gyr old and~${\rm [M/H]}\sim$-0.4][]{seth2010}, indicating
overall similarity between the bulge and disk populations in NGC~404,
partially explaining the lack of radial trends in our SFHs.

\begin{figure}
\centerline{\psfig{file=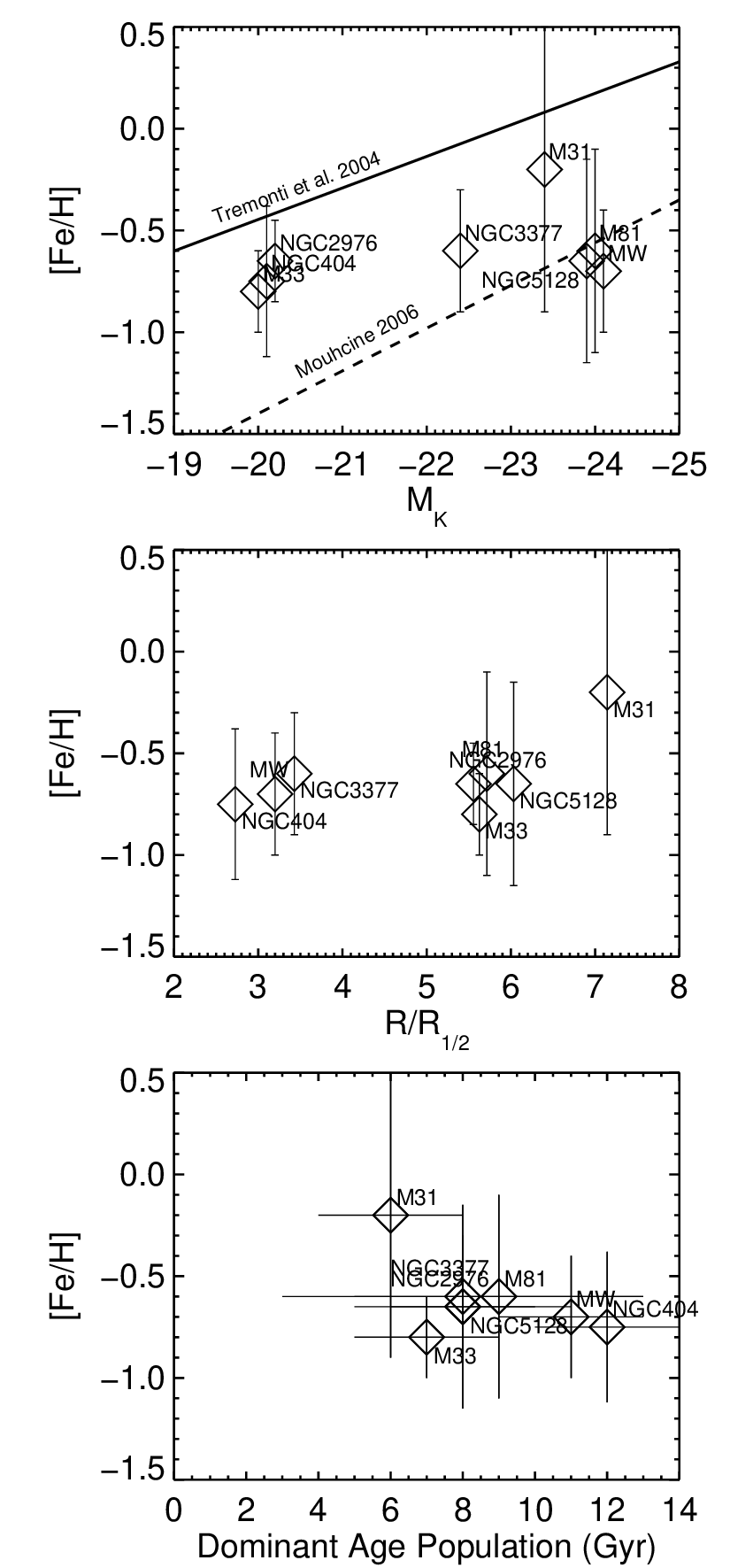,height=5.0in,angle=0}}
\caption{{\footnotesize Metallicity range of the dominant stellar
populations in deep resolved photometry for M31 \citep{brown2006}, M33
\citep{barker2007}, M81 \citep{williams2008}, NGC 5128
\citep{rejkuba2005}, NGC 3377 \citep{harris2007}, NGC~2976
\citep{williams2009b}, and NGC~404 (this work) are plotted along with
that of the Milky Way thick disk \citep{allendeprieto2006} against
several other properties (\S~\ref{ancient}). {\it Top:} The
metallicities as a function of the absolute K-band magnitude of the
galaxy \citep{skrutskie2006}.  The solid and dashed lines show the
luminosity-metallicity relations determined by \citet[][gas phase
metallicity]{tremonti2004} and \citet[][stellar red peak metallicities
of galaxy ``halos'']{mouhcine2006}, respectively.  These relations
were converted from B-band and V-band to K-band using the Tully-Fisher
calibrations of \citet{verheijen2001} and \citet{sakai2000}.  The
Milky Way luminosity was calculated by applying $V_{rot}$= 220 km
s$^{-1}$ to the Tully-Fisher calibration of
\citet{verheijen2001}. {\it Middle:} The populations' metallicities as
a function of the radii at which they were sampled (normalized to the
half-light radius of the galaxy).  {\it Bottom:} The populations'
metallicities as a function of their ages.}}
\label{galaxies}
\end{figure}

The typical age and metallicity of the stars in NGC~404 can be
compared to those of other nearby galaxies, as was done in
\citet{williams2008}.  This comparison is shown in
Figure~\ref{galaxies}.  It shows the NGC~404 disk fields have the
oldest typical age of any disk fields in the nearby sample; however,
the age was determined closer to the galaxy center than for any of the
other galaxies.  Note that the 2 oldest populations, those of the
Galactic thick disk and NGC~404, are those measured closest to the
galaxy centers.

\subsection{The Young Stellar Population}\label{young}

\begin{figure}
\centerline{\psfig{file=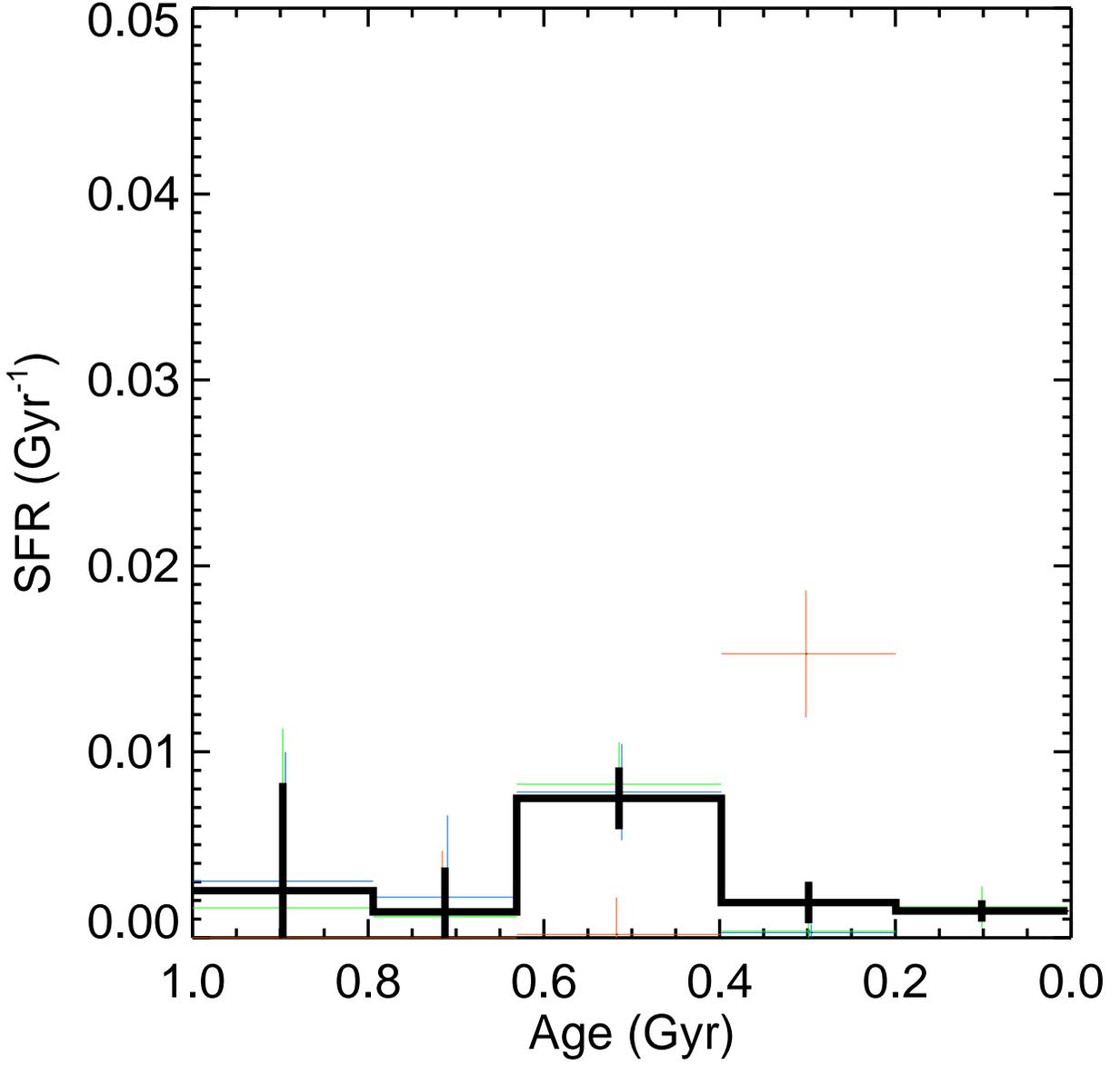,width=6.5in,angle=0}}
\caption{{\it Colored error bars:} The recent star formation histories
of the three fields.  Colors, normalization, and units are the same as
in \ref{sfh_full}. {\it Black histogram:} Combination of all of the
SFHs of the individual regions. The combination shows a significant
increase in the star formation rate $\sim$500~Myr ago
(\S~\ref{young}).}
\label{recent_full}
\end{figure}

Although the star formation in NGC~404 over the past Gyr amounts to an
insignificant fraction of the stellar mass of the disk, it is of
interest for understanding any radial trends in star formation
activity as well as the origin of gas in the NGC~404 disk and its
likely fate.  We show the most recent Gyr of the SFHs of the full
fields in Figure~\ref{recent_full}, along with the weighted average of
all star formation in all fields.  We show these results at a higher
time resolution ($\sim$200~Myr), since our time resolution is good at
these ages due to the rapidly-evolving bright main- and blue He
burning sequences, which contain information for ages $\lap$500 Myr.
For the outer disk fields, we were not able to reliably determine ages
from 400--1000~Myr, due to the small numbers of stars in these fields.
We therefore limited the age range to just two time bins for those
fields.

\begin{figure}
\centerline{\psfig{file=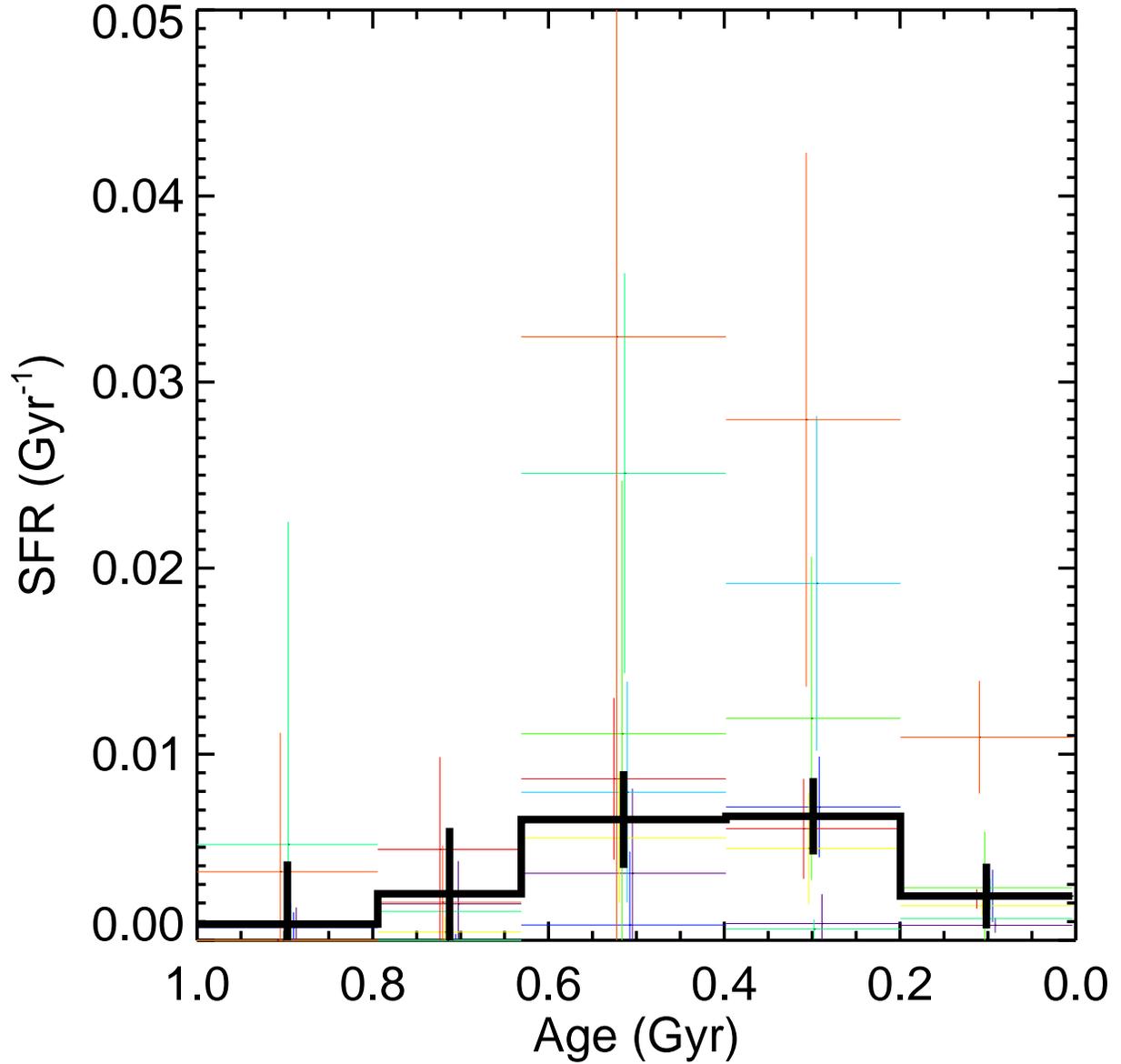,width=6.5in,angle=0}}
\caption{{\it Colored error bars:} The recent star formation histories
of all of the annuli outside of the innermost, normalized by total
stellar mass in the field.  This normalization produces units of
fraction of the total field stellar mass produced each Gyr
(Gyr$^{-1}$). {\it Black histogram:} Combination of all of the SFHs of
the individual regions. The combination shows a significant increase
in the star formation rate $\sim$500~Myr ago (\S~\ref{young}).  The
combined values can be converted from Gyr$^{-1}$ to M$_{\odot}$
yr$^{-1}$ kpc$^{-2}$ by multiplying the Gyr$^{-1}$ values by 0.033.}
\label{recent_ann}
\end{figure}

\begin{figure}
\centerline{\psfig{file=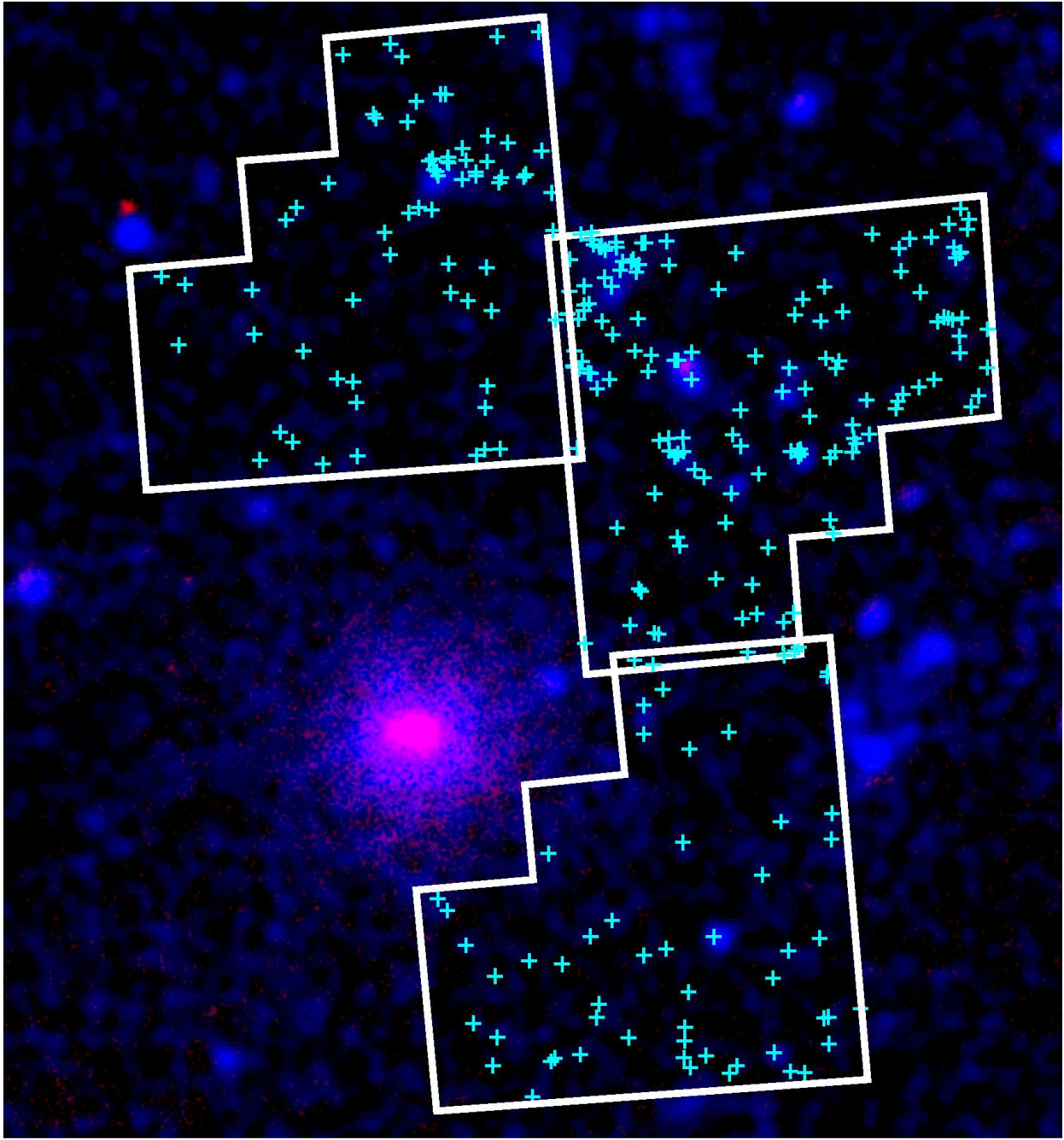,width=6.5in,angle=0}}
\caption{The locations of the upper-main sequence stars from our {\sl
    HST} images (cyan crosses) are shown on a composite GALEX far-UV
    (blue), H$\alpha$ (red) image (\S~\ref{young}).  The FUV-bright
    regions all contain these young stars, but the young stars are not
    confined to the FUV-bright regions. Also note the very low amount
    of H$\alpha$ emission present.}
\label{ms_stars}
\end{figure}

There do not appear to be any radial trends nor any strong positional
dependencies of the recent SFH in the our NGC~404 disk fields (see
discussion below).  However, a distinct pattern emerges in the global
SFHs of the full fields as well as all of the annuli beyond the
innermost (and shallowest) one.  The pattern is plotted in
Figures~\ref{recent_full} and~\ref{recent_ann}, which show all of the
recent SFHs in color and their mean in black.  The best fits to all
fields suggest that star formation increased substantially $\sim$500
Myr ago, rising from consistent with zero to $\sim$2$\times$10$^{-4}$
M$_{\odot}$ yr$^{-1}$ kpc$^{-2}$ by $\sim$400~Myr ago, but then
falling to a lower level by the present.  Indeed, no star formation is
required during the interval 1.3--0.3~Gyr ago to produce
statistically-acceptable fits to the CMD of the deep field; likewise
no star formation is required during the interval 2--0.3~Gyr ago to
fit the NE field and during the interval 1.3--0.2~Gyr ago to fit the
SW field.  Likewise, no star formation more recent than $\sim$100~Myr
is required to fit the CMDs of any of the fields either, showing that
any very recent ($<$100~Myr) star formation has been at a very low
level, consistent with the lack of any diffuse H$\alpha$ emission
(Figure~\ref{ms_stars}).  Thus all of our measurements (outside of the
innermost annulus) {\it require} star formation in the past few
hundred Myr.  While our data allow the possibility that this star
formation was confined to the past 300~Myr, the best fits to our data
have this star formation confined mainly to a small event that
occurred $\sim$600--200~Myr ago.

\begin{figure}
\centerline{\psfig{file=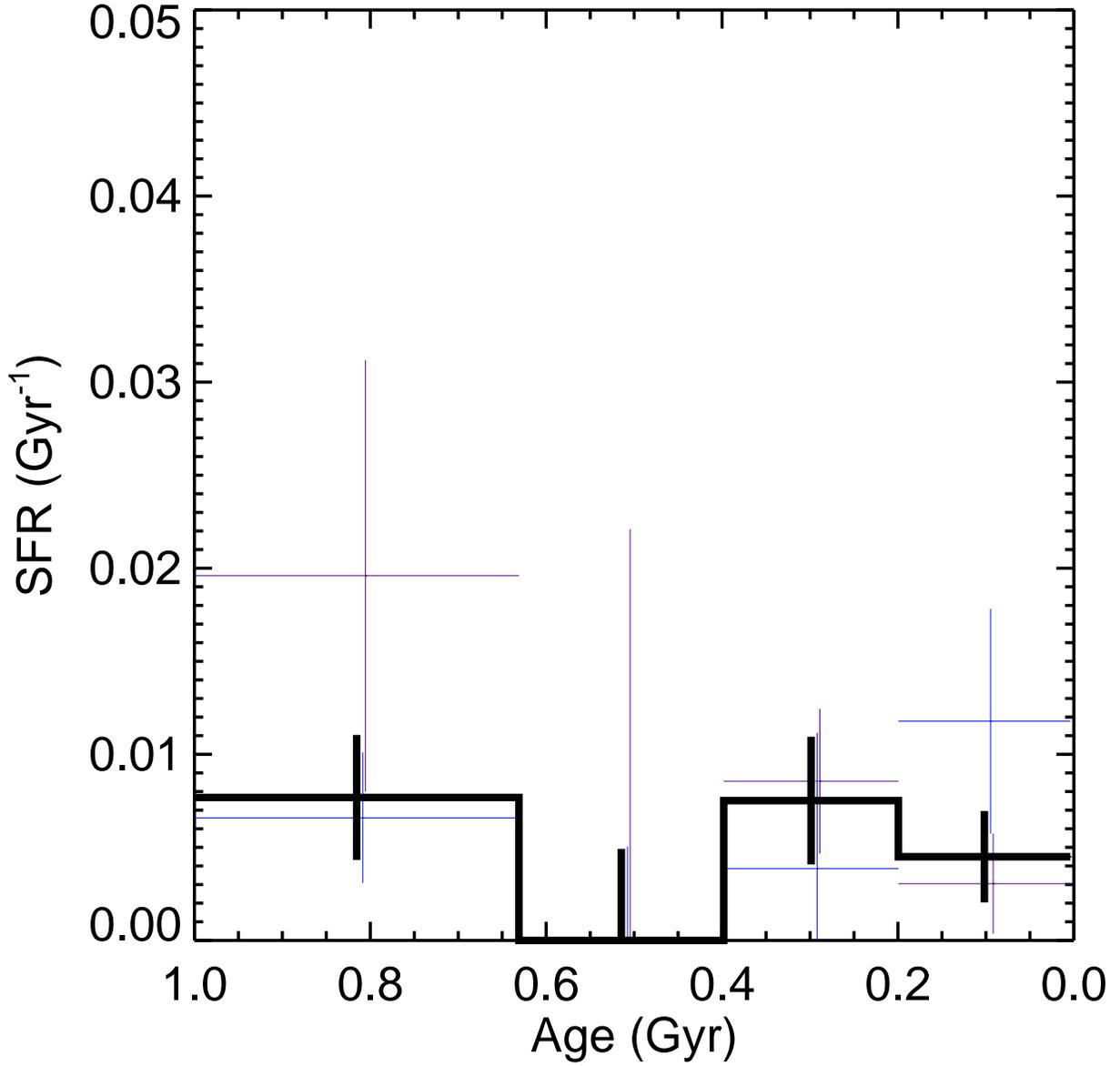,width=6.5in,angle=0}}
\caption{{\it Colored error bars:} The recent star formation histories
  of the two outer disk fields, normalized by total stellar mass in
  the field.  This normalization produces units of fraction of the
  total field stellar mass produced each Gyr (Gyr$^{-1}$). {\it Black
  histogram:} Combination of the SFHs of the two outer disk
  fields. The SFH is consistent with an increase in star formation
  beginning $\sim$400~Myr ago, and the star formation rate over the
  past 400~Myr is consistent with a simple scaling of the rate in the
  inner disk fields with density (\S~\ref{young}).}
\label{outer_sfh}
\end{figure}

The recent SFHs from the outer disk data (archival fields S2 and S3,
Figure~\ref{outer_sfh}) are consistent with those obtained from the
inner disk after scaling by stellar density. However, we were unable
to reliably constrain the age distribution earlier than 400~Myr ago
due to the shallow depth and small number of stars in these fields,
which are located $\sim$2.3 scale lengths
\citep[$r_s$=129.5$''$][]{baggett1998} farther out in the disk.  The
best fits to outer disk fields suggest an increase in star formation
beginning at a similar time ($\sim$400~Myr ago) as seen in the inner
disk fields, with an intensity that is a factor of 10 less than in the
inner disk fields, consistent with a simple density scaling.

To look for variations in the most recent ($\lap$160~Myr) star
formation with position in our inner disk fields, we computed the mass
of stars formed in all 11 of our defined regions (4 annuli each in the
deep and NE fields, and 3 annuli in the SW field) during the past 160
Myr according to our measured SFHs.  This age range probes all O- and
early B-type stars, and thus provides a good baseline for comparison
with the FUV fluxes measured from the GALEX data.

\begin{figure}
\centerline{\psfig{file=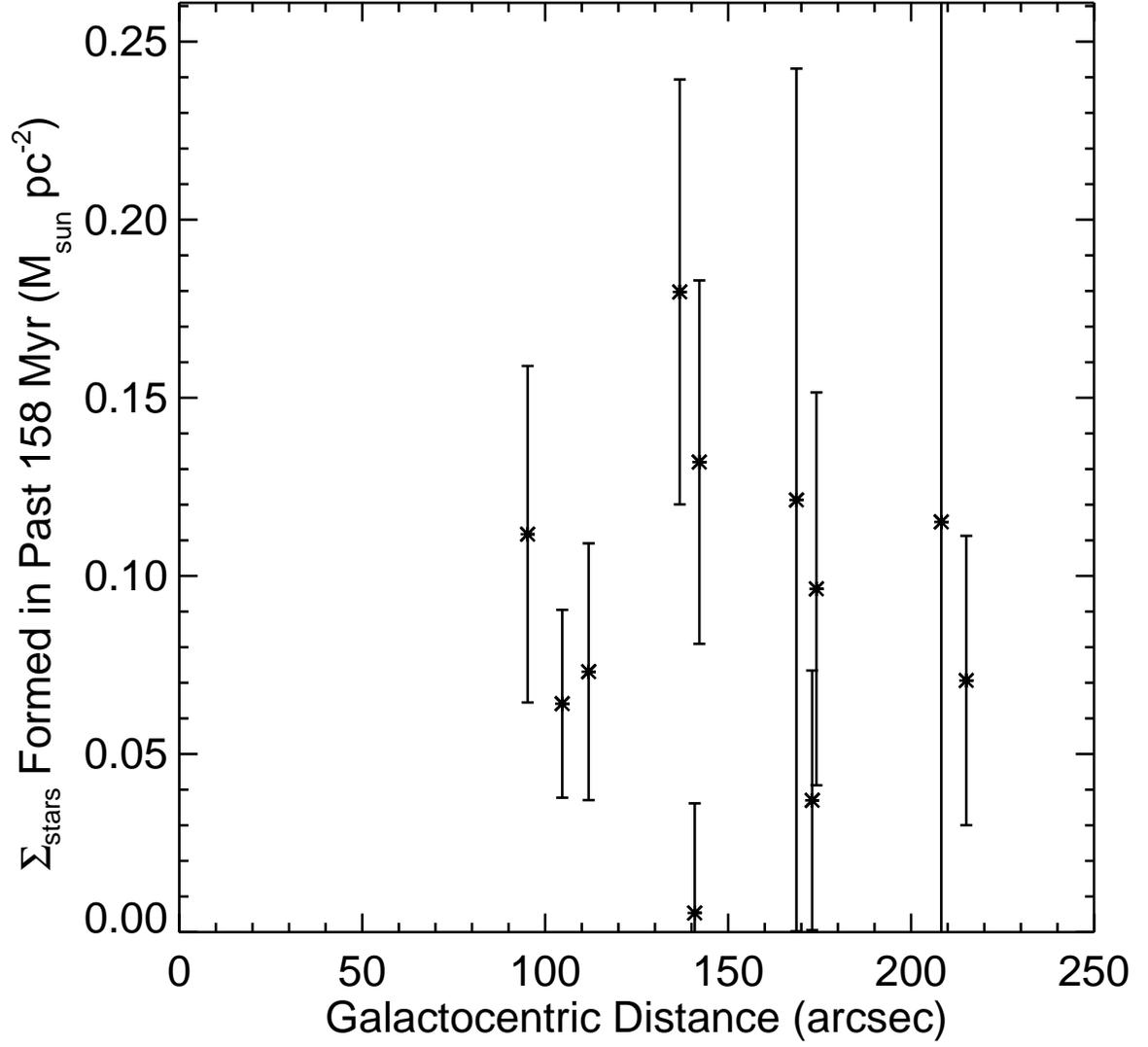,width=6.5in,angle=0}}
\caption{Surface density of stars formed in the past 158 Myr as a
function of galactocentric distance in the NGC~404 disk.  No trend
with radius is apparent (\S~\ref{young}).}
\label{mass_vs_r}
\end{figure}

We plot the median radius of the stars in each of our regions vs. the
mass of stars formed in the past 158~Myr in Figure~\ref{mass_vs_r}. No
radial trend is seen, suggesting that overall the recent SFH of the
disk has not been a strong function of radius in the inner ($<$5~kpc)
disk, and is more consistent with star formation percolating
stochastically across the face of the inner disk
\citep{mcquinn2009}. These results suggest that there is little, if
any, change in the young disk stellar populations with galactocentric
distance in NGC~404 from 1 to 9 arcminutes, which covers $\sim$3.7
scale lengths of the disk and $\sim$9 effective radii.

\begin{figure}
\centerline{\psfig{file=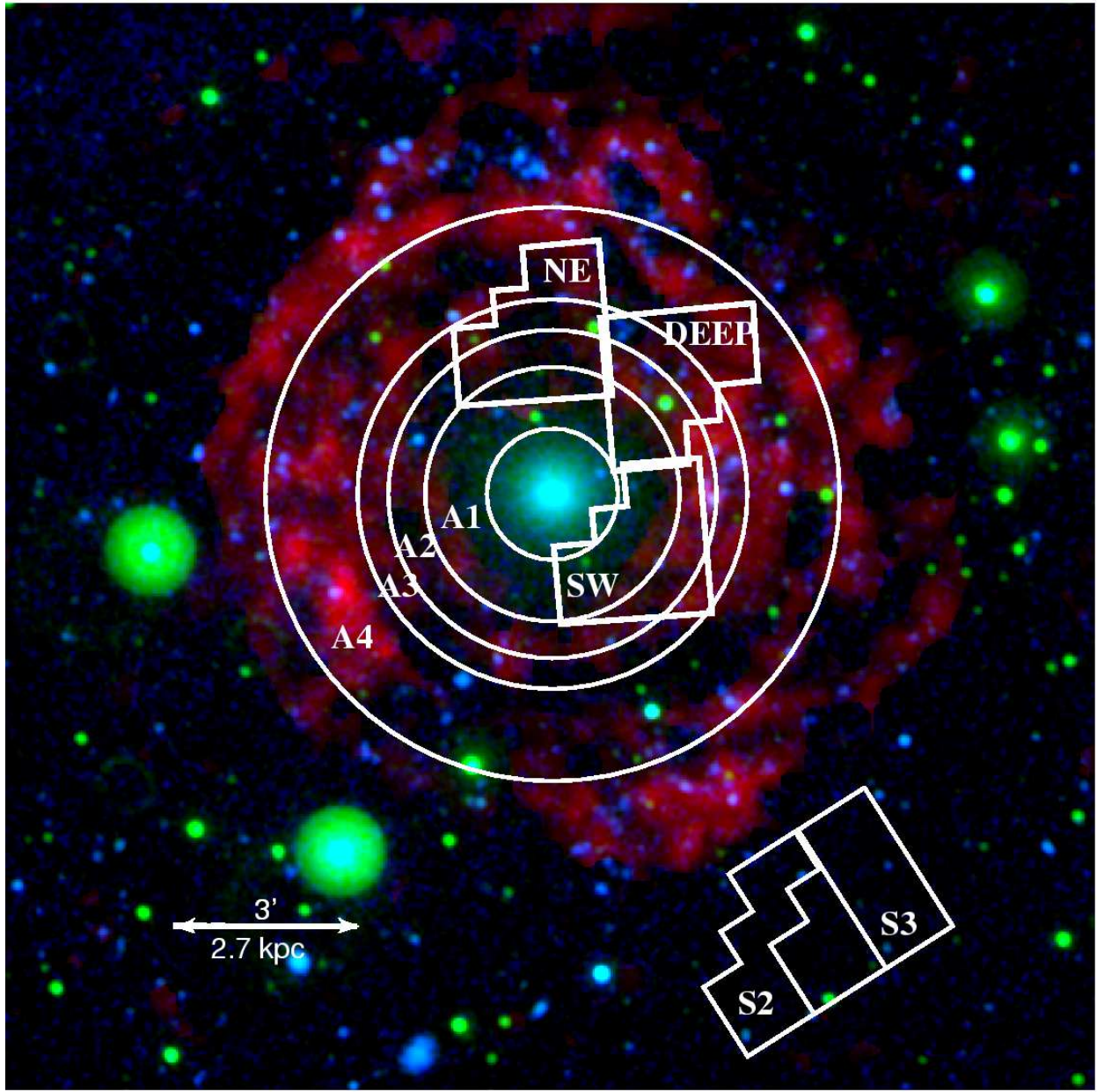,width=6.5in,angle=0}}
\caption{The locations of our NGC~404 fields (\S~\ref{acquisition})
are shown on a 3-color image using GALEX far-UV (blue), GALEX near-UV
(green), and {\sc H i} (red) images. North is up.  East is
left. Fields are labeled with the names used in the text.  The radial
annuli used for the analysis (\S~\ref{division}) are also shown and
labeled.}
\label{field_loc}
\end{figure}

\begin{figure}
\centerline{\psfig{file=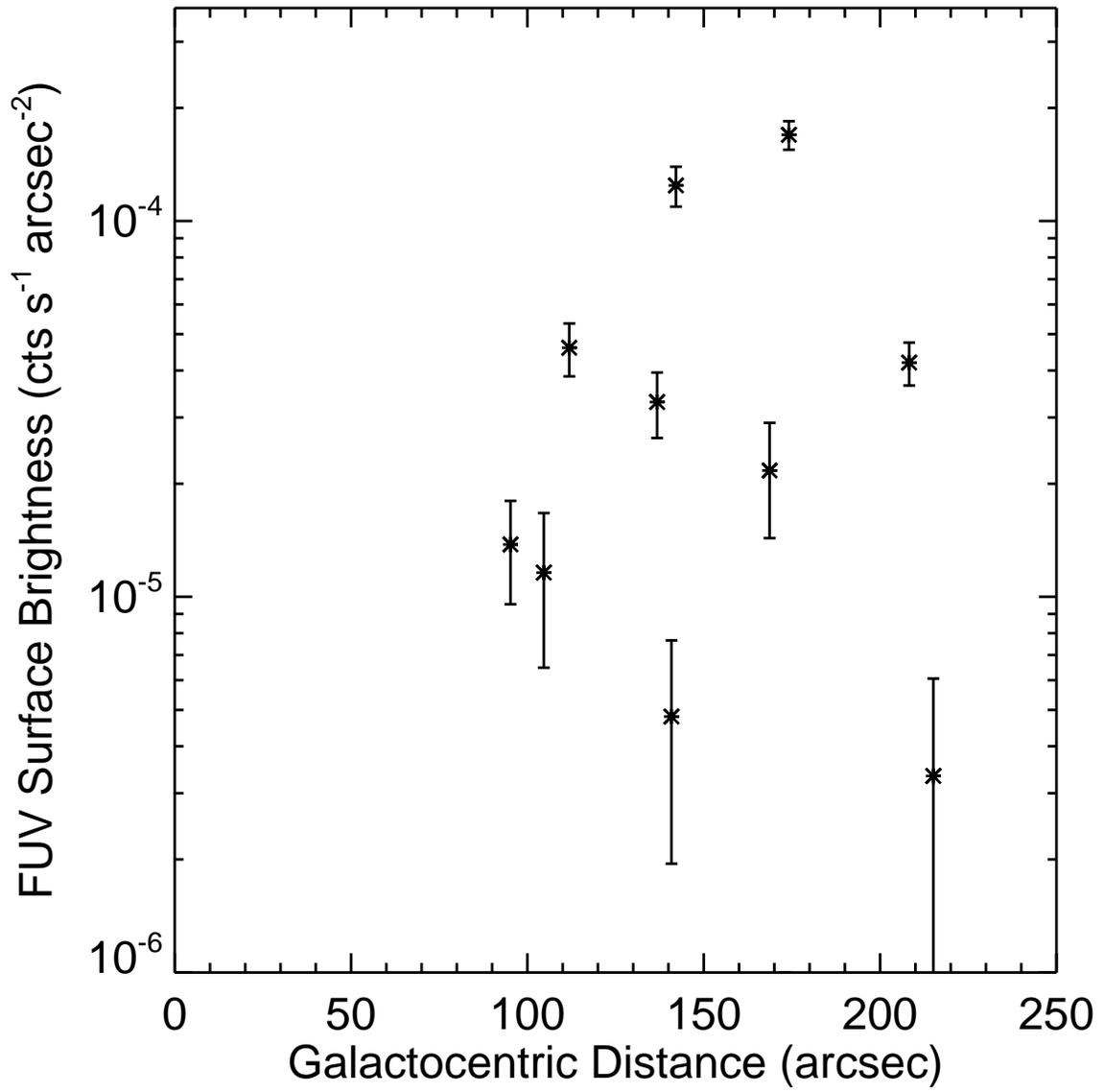,width=6.5in,angle=0}}
\caption{Galex FUV surface brightness as a function of galactocentric
distance in the NGC~404 disk.  No trend with radius is apparent
(\S~\ref{young}).}
\label{fuv_gcd}
\end{figure}

\begin{figure}
\centerline{\psfig{file=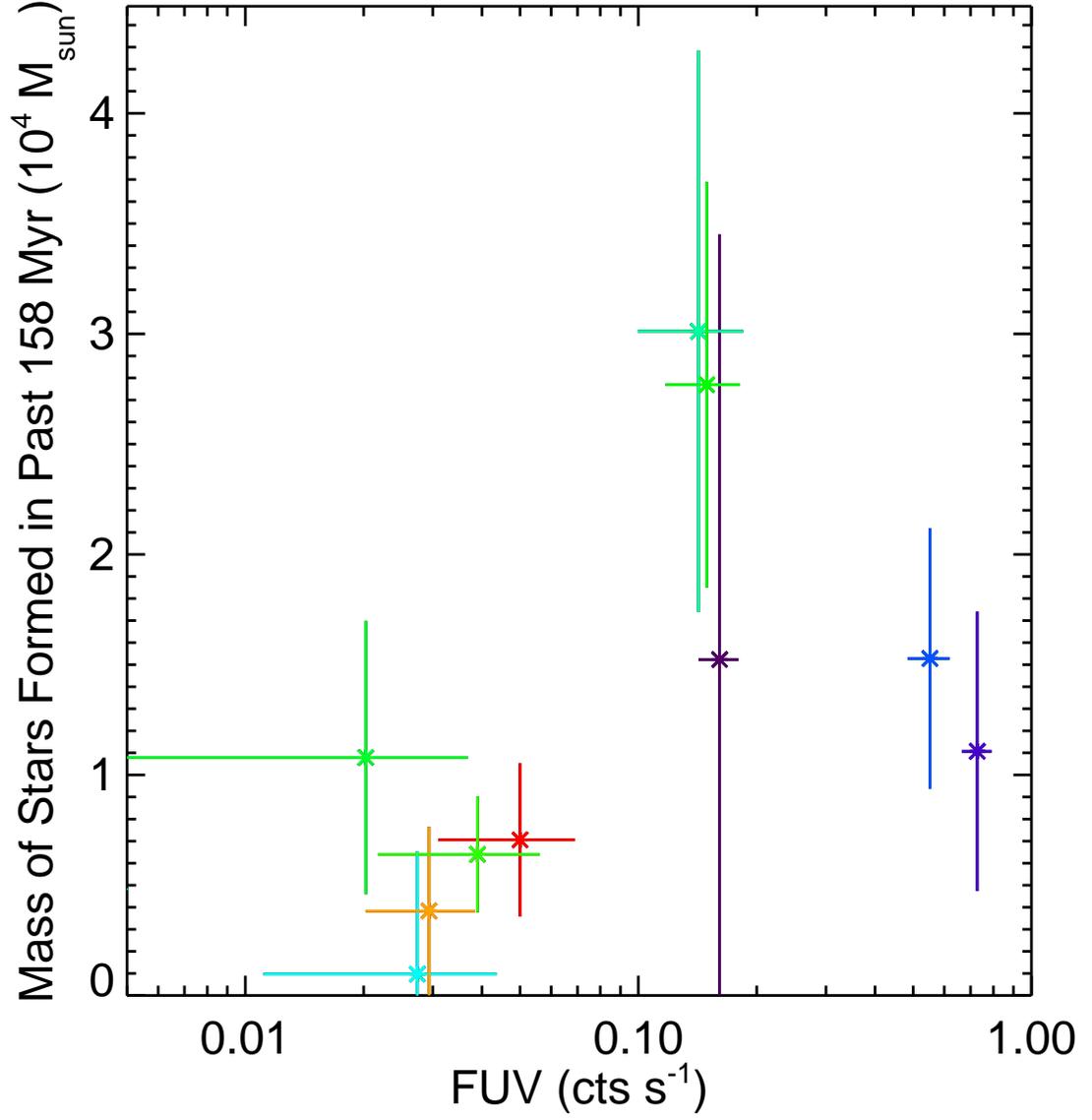,width=6.5in,angle=0}}
\caption{Total stellar mass formed in the past 158 Myr as a function
  of total Galex FUV flux measured in the regions shown in
  Figure~\ref{field_loc}, defined by the WFPC2 field edges and radial
  annuli.  Redder colors denote higher mean ages for the stars formed
  in the past 158 Myr. Only a slightly significant trend is seen,
  partially due to the differences in the mean age of the young
  stellar mass in the different regions (\S~\ref{young}).}
\label{mass_galex}
\end{figure}

We have further tested our ability to detect the star formation
responsible for the flux in the Galex FUV image, which appears as
discrete knots of star formation in the NGC~404 disk (see
Figure~\ref{field_loc}).  We first measured the FUV luminosity of
these knots in each of our defined regions (see
Figure~\ref{field_loc}) using SExtractor \citep{bertin1996}.  We then
subtracted the expected foreground/background luminosity as determined
from a control region 10 scale lengths from the galaxy center.  We
find that the FUV luminosity does not correlate with galactocentric
distance (see Figure~\ref{fuv_gcd}), which is consistent with our
result that the amount of recent ($\lap$160~Myr) star formation does
not correlate with galactocentric distance (other than simply scaling
with density).  On the other hand, there is a hint of a correlation
between the FUV luminosity of the knots and the mass of stars formed
in the past 160 Myr, shown in Figure~\ref{mass_galex}.  A Spearman
Rank test reveals that a correlation is present at only the 2$\sigma$
level (5\% probability of no correlation).

The fact that the correlation is weak may seem surprising, but the
reason for the weakness is apparent when one looks at the mean age of
the stars younger than 160~Myr in each region.  The colors of the
error bars in Figure~\ref{mass_galex} indicate this mean age, with
redder colors representing older ages.  With this additional
information, it is clear that the most FUV-luminous regions contain
the youngest stars, but not the highest stellar masses.  At a given
age (similar color points), the FUV luminosity correlates better with
the stellar mass of young stars.  

Such a scatter in age should be refected in the FUV-NUV colors of the
regions.  We measured these colors and found $\pm$0.8 mag rms scatter.
About 0.3 mag of scatter is expected from photometric errors as
determined by photon statistics.  Some of the additional scatter is
due to the presence of foreground stars in the NUV.  Varying
extinction is not severe enough to explain any additional scatter.  In
the GALEX bandpasses, $A_{FUV}/E_{(B-V)}$=8.376 and
$A_{NUV}/E_{(B-V)}$=8.741 \citep{wyder2005}, or
$E_{(FUV-NUV)}/E_{B-V}$=-0.365.  Thus, the scatter in $E_{(B-V)}$
needed to explain the measured scatter in FUV-NUV is $\pm$2.2 mag.
Such a large amount of differential extinction would be far more than
we observe in our CMDs.  While we cannot easily quantify the effects
of foreground contamination, the scatter in FUV-NUV color supports the
idea that a range of ages is in large part responsible for the
weakness of the correlation.

Furthermore, the distribution of upper-main sequence stars on the
GALEX FUV image, shown in Figure~\ref{ms_stars}, suggests that these
young stars are more widely distributed than the FUV emission. While
all FUV-bright regions contain upper-main sequence stars, Not all
upper-main sequence stars are confined within FUV-bright regions.  The
different distributions could be explained by stars migrating away
from their birth regions on timescales $<$160~Myr \citep[consistent
with][]{lada2003}, and FUV emission being dominated by stars younger
than 100~Myr \citep[consistent with][]{gogarten2009}.  These results
show that there is a complex relationship between UV properties, age,
and stellar mass in regions containing recent star formation.

\subsection{The Origin of the Gas Disk}\label{gasdisk}

It is interesting that our SFHs show an increase in star formation
$\sim$0.5 Myr ago, in light of the suggestion of \citet{delrio2004}
that the gas disk may have originated in a merger event with a dIrr
galaxy 0.5--1 Gyr ago.  However, our full-field CMDs all require some
star formation between 2 and 10 Gyr, as well as star formation between
0.5 and 0.1 Gyr ago, to produce an acceptable fit to the observed CMD.
Failure to include stars in the 2--10 Gyr age range produces a model
CMD containing more blue AGB stars above the RGB tip than are seen in
the data.  These stars produce a quantitatively significant decrease
in the quality of the model fit.  The lower limits on the star
formation rates in the 2--10 Gyr period are comparable to the peak
rates over the past 500~Myr.

To investigate the origin of the gas in the disk, we studied the
correlation between gas density and star formation in the disk of
NGC~404 over its history. We assume that all of the stars that formed
during a given time interval in our SFH must have been in the form of
gas at the beginning of that time interval.  When this gas is added to
the gas seen at the present day, we can infer the gas density in the
past, and then compare it to the subsequent star formation rate in our
SFH.  We can do this at increasing lookback times, giving us a way of
probing the correlation between gas density and star formation rate at
a range of epochs.

We begin with the current {\sc H i} column density of the disk from
 2--5$'$ \citep[$\sim$1.5$\times$10$^{20}$~cm$^{-2}$;][]{delrio2004},
 which corresponds to 1.2~M$_{\odot}$~pc$^{-2}$.  We then assume that
 $\Sigma_{\rm H} = \Sigma_{\rm HI}$ and adopt the conversion
 $\Sigma_{gas} = 1.45\Sigma_{\rm H}$ \citep{kennicutt1989} to account
 for metals and molecular gas.  Finally, we assume that all stars
 formed from a previously-existing gas disk that was in place at the
 beginning of each time interval.  Our method therefore requires that
 the gas disk had a higher surface density in the past in order to
 provide the material for all of the star formation that subsequently
 occurred. We further assume that stellar evolution quickly recycles
 20\% of the mass of stellar mass back to the gas phase.  This
 fraction corresponds to all stars $>$8~M$_{\odot}$ (assuming a Kroupa
 IMF integrated from 0.1 M$_{\odot}$ to 100 M$_{\odot}$), which return
 their mass to the ISM on timescales shorter than 100 Myr.  The exact
 value of this fraction had minimal impact on our estimates.

\begin{figure}
\centerline{\psfig{file=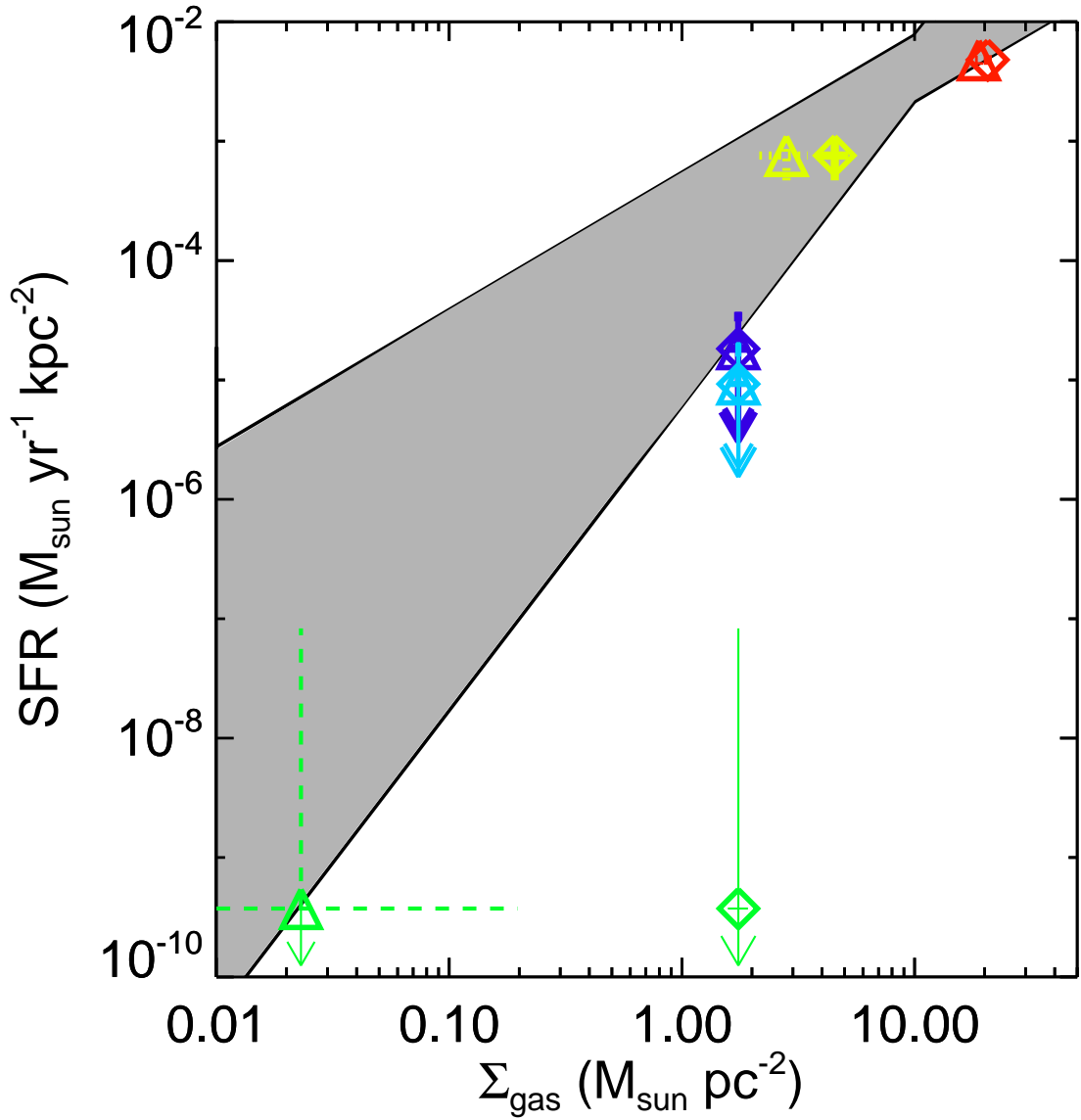,width=6.5in,angle=0}}
\caption{{\it Diamonds with solid errors:} Gas surface density at
previous epochs assuming all gas was in place 14 Gyr ago.  Three
epochs are plotted.  Redder colors denote older ages (0.3,0.6,0.9,5,12
Gyr).  {\it Triangles with dotted errors:} Gas surface density at
previous epochs assuming a gas accretion event occurred 0.6--0.9 Gyr
ago.  Colors are the same ages as the diamonds.  {\it Shaded Area:}
The regions of gas density - star formation rate density space covered
by the relation measured by THINGS \citep{bigiel2008}.  Note that the
star formation rate was far too low for the assumed gas density
0.9~Gyr ago if we assume no gas has been accreted (light blue diamond;
\S~\ref{gasdisk}).}
\label{schmidt}
\end{figure}

 With the above assumptions, we were able to infer the gas density of
 the disk for several epochs in the history of NGC~404 and compare
 that density to our measured star formation rates.  We were then able
 to test the plausibility of the assumptions by comparing the results
 to the known correlation between gas density and star formation rate
 \citep{bigiel2008}. The results of this reconstruction of the history
 of the gas density and star formation rate in the NGC~404 inner disk
 are shown in Figure~\ref{schmidt}, where diamonds show the gas
 density and mean star formation rates of the disk for each of 5
 epochs (0.3, 0.6, 0.9, 5, 12 Gyr ago) calculated from our measured
 SFHs.  The $\Sigma_{\rm SFR}$ vs.  $\Sigma_{\rm gas}$ in the early
 epochs agrees with the overall trends measured by the {\sc H i}
 Nearby Galaxy Survey (THINGS) \citep{bigiel2008} under the assumption
 that all of the stars formed from a gas disk that was formed
 $\sim$14~Gyr ago.  However, at 0.9 Gyr, the star formation rate is
 inconsistent with the gas density under this ``closed box''
 assumption.

To test the possibility that a recent merger with a small gas-rich
galaxy may have affected the evolution of the NGC~404 disk, we
performed a reconstruction of the past gas density and star formation
rate described above, but we altered the closed-box assumption.
Instead, we assume that 10$^8$~M$_{\odot}$ of gas was added to the
system 0.6--0.9 Gyr ago.  The effect of this assumption is to make the
gas density of the disk significantly lower $\sim$1 Gyr ago than it is
today, unlike the closed box assumption where the gas density strictly
increases with lookback time.  The triangle points with dotted errors
in Figure~\ref{schmidt} show the results of this reconstruction.
These points, including the one at 0.9 Gyr, are all consistent with
the known correlation between gas density and star formation rate.
This consistency allows the possibility that gas was recently
accreted, as was suggested by \citet{delrio2004} based on the warped
nature of the outer gas disk.

Together, these results favor a scenario where a gas disk was in place
for many Gyr.  This disk was of low surface density
($<$5~M$_{\odot}$~pc$^{-2}$) for most of the history of the disk and
formed stars at a low rate
($<$10$^{-3}$~M$_{\odot}$~yr$^{-1}~$kpc$^{-2}$).  This gas may have
been completely depleted by $\sim$1 Gyr ago, when new gas entered the
system, either from surrounding gas filaments, or from a small,
gas-rich, dwarf galaxy that merged to produce the current gas disk.
This event triggered a small amount of new star formation over the
past $\sim$500~Myr.

It is still possible that the present gas disk is a relic of the gas
that originally formed the disk and that there was no recent merger.
If so, the low-density gas disk must have formed stars in short, weak,
episodes that lasted $\sim$200--400 Myr.  Such episodes, like the one
we have resolved in the most recent Gyr, form only $\lap$10$^7$
M$_{\odot}$ of stellar mass in the inner disk ($r\,<\,5$~kpc, assuming
axisymmetry), which could reconcile the amount of stars formed in the
past Gyr with the amount of gas available in our closed-box
assumption.  Such star formation would be insignificant compared to
the $\gap$2$\times$10$^9$ M$_{\odot}$ of stellar mass formed in the
inner disk prior to 10~Gyr ago and the 1.5$\times$10$^8$ M$_{\odot}$
of {\sc H i} present \citep{delrio2004}.  Such a scenario is favored
by the isolated environment of NGC~404 but is not consistent with the
star formation rates we measure from 0.6--1.0 Gyr.  Furthermore,
\citet{delrio2004} suggest the warped nature of the gas disk as
additional evidence for the recent merger.  While the warp is
consistent with the merger scenario, warping can also occur from
misalignment of angular momentum during late gas accretion from
filaments \citep{binney1999,shen2006} and therefore does not require a
catastrophic merger.

\section{Conclusions}

We have measured resolved stellar photometry for $\sim$15 arcmin$^2$
of the inner disk and $\sim$10 arcmin$^2$ of the outer disk of the S0
galaxy NGC~404.  Detailed fitting of the resulting CMDs shows that the
disk is dominated by stars older than $\sim$10~Gyr, with all of our
inner disk data constraining the age of $\sim$90\% of the stellar mass
to $>$8~Gyr and our best data constraining the age of 75\% of the
stellar mass to $>$10~Gyr.  These results show that the difference
between this disk and most later-type disks that have been studied in
detail is seen in the old stellar population as well as the young.

We found no trends between the age distributions of the young or old
populations and distance from the galaxy center within the observed
regions.  However, as seen by \citet{tikhonov2003} the mean RGB color
and AGB/RGB ratio of the innermost regions are consistent with a bulge
population in NGC~404 that is younger than the disk.

Despite its low gas density, NGC~404 appears to have been forming
stars throughout its history on timescales of Gyr, even in the inner
disk ($\sim$1--1.5$'$).  The SFH and the current density of the gas
disk are consistent with NGC~404 having evolved roughly in a closed
box over much of its history, starting as a large disk of high surface
density and approximately following the Schmidt law as its overall
star formation rate and gas density decreased.  Such long-lived disks
have been observed in isolated galaxies before, as in the compact
dwarf ADBS 113845+2008 \citep{cannon2009}, whose gas disk is also of
low density.  Perhaps these low density disks have very long lifetimes
due to very low star formation efficiencies at these densities.  Our
SFHs indicate that star formation in the low density relic disk may be
episodic, going through minor episodes of star formation that last a
few hundred Myr every Gyr or so.

On the other hand, in NGC~404 there is some evidence that this passive
evolution was slightly disrupted by an event 0.6-0.9 Gyr ago, possibly
a small merger.  Such a merger event would alleviate the necessity for
very low star formation efficiency $\sim$1~Gyr ago by allowing that
the currently-observed gas disk was not fully in place at that time.

Support for this work was provided by NASA through grants GO-10915 and
GO-11719 from the Space Telescope Science Institute, which is operated
by the Association of Universities for Research in Astronomy,
Incorporated, under NASA contract NAS5-26555.

\clearpage

\begin{deluxetable}{lccccc}
\tablecaption{Properties of the Designated Regions}
\tablehead{
\colhead{{\footnotesize Region}} &
\colhead{{\footnotesize R$_{in}$ ($''$)}} &
\colhead{{\footnotesize R$_{out}$ ($''$)}} &
\colhead{{\footnotesize R$_{med}$ ($''$)\tablenotemark{a}}} &
\colhead{{\footnotesize $F606W_{50}$\tablenotemark{b}}} &
\colhead{{\footnotesize $F814W_{50}$\tablenotemark{c}}}
}
\startdata
{\footnotesize deep annulus 1} & 64.4 & 124.0 & 104.7 & 26.6 & 26.0\\
{\footnotesize deep annulus 2} & 124.0 & 160.0 & 142.1 & 27.3 & 26.3\\
{\footnotesize deep annulus 3} & 160.0 & 190.0 & 174.1 & 27.8 & 26.9\\
{\footnotesize deep annulus 4} & 190.0 & 268.4 & 215.1 & 28.0 & 27.2\\
{\footnotesize NE annulus 1} & 93.0 & 124.0 & 111.9 & 26.8 & 25.8\\
{\footnotesize NE annulus 2} & 124.0 & 160.0 & 140.9 & 26.9 & 26.1\\
{\footnotesize NE annulus 3} & 160.0 & 190.0 & 173.0 & 27.0 & 26.1\\
{\footnotesize NE annulus 4} & 190.0 & 250.5 & 208.3 & 27.2 & 26.3\\
{\footnotesize SW annulus 1} & 64.0 & 124.0 & 95.2 & 26.6 & 25.8\\
{\footnotesize SW annulus 2} & 124.0 & 160.0 & 136.8 & 26.9 & 26.0\\
{\footnotesize SW annulus 3} & 160.0 & 190.0 & 168.7 & 26.9 & 26.2\\
{\footnotesize S2} & 404.8 & 565.2 & 468.3 & 26.8 & 26.1\\
{\footnotesize S3} & 406.7 & 560.6 & 480.8 & 26.9 & 26.4\\
\enddata
\tablenotetext{a}{The median galactocentric distance of the stars in the the region.}
\tablenotetext{b}{The 50\% completeness limit of the F606W data.}
\tablenotetext{c}{The 50\% completeness limit of the F814W data.}
\label{table}
\end{deluxetable}

\clearpage

\end{document}